\documentclass[prd,aps,twocolumn,showpacs,floats,
letterpaper,floatfix,superscriptaddress]{revtex4}
\usepackage[dvips]{graphicx}

\newcommand{\eqref}[1]{(\ref{#1})}

\begin{document}

%\fnsymbol{footnote}

\title{Numerical evolutions of nonlinear $r$-modes in neutron stars}

\author{Lee Lindblom}
\affiliation{Theoretical Astrophysics 130-33, California Institute of Technology, Pasadena, CA 91125}
\author{Joel E. Tohline}
\affiliation{Department of Physics and Astronomy, Louisiana State University, Baton Rouge, LA 70803}
\author{Michele Vallisneri}
\affiliation{Theoretical Astrophysics 130-33, California Institute of Technology, Pasadena, CA 91125}

\date{November 30, 2001}

\begin{abstract}
Nonlinear evolution of the gravitational radiation (GR) driven
instability in the $r$-modes of neutron stars is studied by full
numerical 3D hydrodynamical simulations.  The growth of the $r$-mode
instability is found to be limited by the formation of shocks and
breaking waves when the dimensionless amplitude of the mode grows to
about three in value.  This maximum mode amplitude is shown by
numerical tests to be rather insensitive to the strength of the GR
driving force.  Upper limits on the strengths of possible nonlinear
mode--mode coupling are inferred.  Previously unpublished details
of the numerical techniques used are presented, and the results of
numerous calibration runs are discussed.
\end{abstract}

\maketitle

\section{\label{intro}Introduction}

In recent years the gravitational radiation (GR) driven instability in
the $r$-modes of rotating neutron stars has received considerable
interest, both as a source of gravitational waves for detectors such
as the Laser Interferometer Gravitational Wave Observatory (LIGO), and as an astrophysical process capable of limiting the rotation rates of neutron stars.  In any rotating star, the $r$-modes
are driven towards instability by GR \cite{Andersson98,Friedman98}: as
the star emits gravity waves (primarily through a gravitomagnetic
effect), the GR reaction acts back on the fluid by lowering the
(already negative) angular momentum of the mode.  This in turn causes
the amplitude of the mode to grow.  In most stars internal dissipation
suppresses the $r$-mode instability, but this may not be the case for
hot, rapidly rotating neutron stars \cite{Lindblom98,Lindblom01b}. For
neutron stars with millisecond rotation periods, the timescale for the
growth of the instability is about 40 s. In the absence of any
limiting process, GR would force the dimensionless amplitude of the
most unstable ($m=2$) $r$-mode to grow to a value of order unity
within about ten minutes of the birth of such a star. (At unit
amplitude, the characteristic $r$-mode velocities are comparable to
the rotational velocity of the star.)

The strength of the GR emitted and the timescale on which the neutron
star loses angular momentum and spins down depend critically on the
maximum amplitude to which the $r$-mode grows. Initial estimates
assumed that the amplitude would grow to a value of order unity before
an undescribed nonlinear process saturated the mode. After saturation,
it was assumed that the spindown would proceed as a quasi-stationary
process, reducing the angular velocity to one tenth of its initial
value within about one year. In this scenario, gravitational waves
from spindown events might be detectable with LIGO II \cite{Owen98}.

However, at present no one knows with certainty how large the
amplitude of the $r$-modes will grow. It may well be that the nonlinear
hydrodynamics of the star might limit the growth of $r$-modes to very
small values.  This could happen, for instance, if the $r$-modes were
to leak energy by nonlinear coupling into other modes faster than GR
reaction could restore it. In this case the $r$-mode instability would
not play any interesting role in real astrophysical systems.

In a previous paper \cite{Lindblom01}, we presented the preliminary
results of fully nonlinear, three-dimensional numerical simulations
aimed at investigating the growth of $r$-modes. In our simulations, we
modeled a young neutron star as a rapidly rotating, isentropic,
Newtonian polytrope; we added a small-amplitude seed $r$-mode and we
solved the hydrodynamic equations driven by an effective GR reaction
force. We found that $r$-mode saturation intervenes at amplitudes far
larger than expected ($\sim 3$), supporting the astrophysical
relevance of $r$-modes and the possibility of detecting $r$-mode
gravity waves.  The details of the GR signature emitted by the
$r$-mode instability that we observe in our simulations are rather
different than previously envisioned, and these details suggest that
this radiation may be more easily detected than previously thought:
the radiation is more monochromatic and is emitted in a shorter, more
powerful burst (see Ref.~\cite{Lindblom01} and the final Section of
this paper).

Our results are compatible with the conclusions of Stergioulas and
Font~\cite{Stergioulas01}, who performed relativistic simulations of
$r$-modes on a fixed neutron-star geometry, and found no saturation
even at large amplitudes. A second point of comparison can be made
with the work of Schenk and colleagues \cite{Schenk01}.  They have
attacked the problem analytically, developing a perturbative formalism
to study the nonlinear interactions of the modes of rotating stars,
and proving that the couplings of $r$-modes to many other rotational
modes are small (they are forbidden by selection rules, or they vanish
to zeroth order in the angular velocity of the star).

The present work is meant as a complement to Ref.~\cite{Lindblom01}:
throughout these pages, we describe our simulations in greater detail;
we discuss their relevance and their limitations in the light of
Refs.~\cite{Stergioulas01,Schenk01}; and we present the results of
several additional simulations aimed at enlightening particular
aspects of the problem. In Sec.~\ref{equations} we write down the
basic hydrodynamic equations, and we define a number of mathematical
quantities that will be used to monitor the nonlinear evolution of the
$r$-modes.  In Sec.~\ref{reaction} we implement the effective
current-quadrupole gravitational radiation-reaction force. In
Sec.~\ref{calibration} we integrate the fluid equations with $r$-mode
initial data in slowly rotating stars, and we compare the results with
the small-amplitude, slow-rotation analytical expressions: we
demonstrate that the integration reproduces faithfully the analytical
predictions to the expected degree of accuracy. In
Secs.~\ref{sec:fast}--\ref{sec:k1000b} we study the nonlinear
evolution of $r$-mode initial data in rapidly rotating stars,
concentrating on the nonlinear saturation of the $r$-modes, and
analyzing in detail the evolution of several hydrodynamical
quantities. Finally, we summarize our conclusions in
Sec.~\ref{conclusion}.

\section{\label{equations}Basic Hydrodynamics}

We study the solutions to the Newtonian fluid equations,
\begin{equation}
\partial_t\rho+\vec{\nabla}\cdot\bigl(\rho\vec{v}\bigr)=0,
\label{eq:continuity}
\end{equation}
\begin{equation}
\rho\bigl(\partial_t\vec{v}+\vec{v}\cdot\vec{\nabla}\vec{v}\bigr) =
-\vec{\nabla}p - \rho\vec{\nabla}\Phi + \rho \vec{F}^{\mathrm{GR}},
\label{eq:euler}
\end{equation}
\begin{equation}
\partial_t \tau +\vec{\nabla}\cdot\bigl(\tau \vec{v}\bigr)=0,
\label{eq:energyeq}
\end{equation}
where $\vec{v}$ is the fluid velocity, $\rho$ and $p$ are the
density and pressure, $\Phi$ is the Newtonian gravitational
potential, and $\vec{F}^{\mathrm{GR}}$ is the gravitational
radiation reaction force. Equation \eqref{eq:energyeq} is a
recasting of the energy equation for adiabatic flows, where $\tau$
is the \emph{entropy tracer} \cite{Tohline01}; for polytropic
equations of state, $\tau$ is related to the internal
energy (per unit mass) $\epsilon$ by the relation $\tau = (\epsilon
\rho)^{1/\gamma}$, where $\gamma$ is the adiabatic exponent. The
Newtonian gravitational potential is determined by Poisson's
equation,
\begin{equation}
\nabla^2 \Phi = 4\pi G\rho,
\label{eq:poisson}
\end{equation}
while the gravitational radiation reaction force will be
discussed in Sec.~\ref{reaction}.

We solve Eqs.~\eqref{eq:continuity}--\eqref{eq:poisson}
numerically in a rotating reference frame, using the computational
algorithm developed at LSU to study a variety of astrophysical
hydrodynamic problems~\cite{Tohline97}.  The code performs an
explicit time integration of the equations using a
finite-difference technique that is accurate to second order both
in space and time, and uses techniques very similar to those of
the familiar ZEUS code~\cite{Stone92}. For most of our simulations, we adopt a cylindrical grid with 64 cells in the radial direction, and 128 cells in the axial and azimuthal directions.

In the limit of slow rotation, we define the $r$-modes of rotating
Newtonian stars (using the normalization of Lindblom, Owen and Morsink
\cite{Lindblom98}) as the solutions of the perturbed fluid equations
having the Eulerian velocity perturbation
\begin{equation}
\delta \vec{v} = \alpha_0 R \Omega_0 \Bigl( \frac{r}{R} \Bigr)^l \vec{Y}^{B}_{ll} e^{i \omega_0 t},
\label{eq:rmodedef}
\end{equation}
where $R$ and $\Omega_0$ are the radius and angular velocity of the
unperturbed star, $\alpha_0$ is the dimensionless $r$-mode
amplitude, and $\vec{Y}^{B}_{ll}$ is a vector spherical harmonic
of the magnetic type, defined by
\begin{equation}
\vec{Y}^{B}_{lm} = [l(l+1)]^{-1/2} r \vec{\nabla} \times (r \vec{\nabla} Y_{lm}).
\end{equation}
The $r$-mode frequency is given by \cite{Papaloizou78}
\begin{equation}
\omega_0 = - \frac{(l-1)(l+2)}{l+1} \Omega_0.
\label{eq:omega}
\end{equation}

To monitor the nonlinear evolution of the $r$-modes, it is helpful
to introduce nonlinear generalizations of the amplitude and
frequency of the mode.  These quantities are defined most
conveniently in terms of the current multipole moments of the
fluid,
\begin{equation}
J_{lm} = \int \rho r^{l} \vec{v}\cdot\vec{Y}^{B*}_{lm} d^3 x.
\label{eq:j22}
\end{equation}
In slowly rotating stars, the $J_{22}$ moment is proportional to
the amplitude of the $m=2$ $r$-mode, the most unstable mode, and
the one that we will study. To track the evolution of this mode
even in the nonlinear regime, we define the normalized,
dimensionless amplitude
\begin{equation}
\alpha = \frac{2|J_{22}|}{\tilde{J}MR^3\Omega_0},
\label{eq:alphadef}
\end{equation}
where $M$ is the total mass of the star and $\tilde{J}$ is defined
by
\begin{equation}
\tilde{J} M R^4 = \frac{1}{4\pi} \int \rho r^4 d^3 x \simeq \int
\rho r^6 dr. \label{eq:tildej}
\end{equation}
The quantity $\tilde{J}$ is evaluated once and for all at the
beginning of each of our evolutions. For slowly rotating stars, the
definition \eqref{eq:alphadef} of the mode amplitude reduces to
the one given by Eq.~\eqref{eq:rmodedef}.

In slowly rotating stars, and in all situations where the leading
contribution to $J_{22}$ comes from the $m=2$ $r$-mode, the time
derivative $dJ_{22}/dt$ is proportional to the frequency of the
mode: $d{J}_{22}/dt = i\omega J_{22}$.  Thus we are led to define
the nonlinear generalization of the $r$-mode frequency as
\begin{equation}
\omega = - \frac{1}{|J_{22}|}\left|\frac{d{J}_{22}}{dt}\right|.
\label{eq:genomega}
\end{equation}
As shown by Rezzolla, \emph{et al.}\ \cite{Rezzolla99}, we can
re-express $dJ_{22}/dt$ as an integral over the standard fluid
variables,
\begin{equation}
J^{(1)}_{22} \equiv \frac{d{J}_{22}}{dt} = \int\rho\Bigl[
\vec{v}\cdot\bigl(\vec{\nabla}\vec{Y}^{B*}_{22}\bigr)\cdot\vec{v}
- \vec{\nabla}\Phi\cdot \vec{Y}^{B*}_{22}\Bigr] d^3 x.
\label{eq:dj22}
\end{equation}
The definitions, Eqs. \eqref{eq:alphadef} and \eqref{eq:genomega}, of
mode amplitude and mode frequency are very stable numerically, because
they are expressed in terms of integrals over the fluid variables. In
the Appendix, we give explicit expressions for $J_{22}$ and
$J^{(1)}_{22}$ in the cylindrical coordinate system used in our
numerical analysis.

While we monitor the nonlinear evolution of the $r$-mode, we are
also interested in tracking the star's average angular
velocity as well as its degree of differential rotation. With this
in mind, we define the average angular velocity
%%$\bar{\Omega}$ of the star as the ratio of the angular momentum
%%to the moment of inertia of the star,
%
\begin{equation}
\bar{\Omega} \equiv J/I, \label{eq:baromega}
\end{equation}
where the angular momentum and the moment of inertia are given 
respectively by
\begin{equation}
J = \int \rho \varpi^2 \Omega(\varpi,z,\varphi) d^{3} x,
\label{eq:jint}
\end{equation}
\begin{equation}
I = \int \rho \varpi^2 d^3 x.
\label{eq:iint}
\end{equation}
Here $\varpi$ is the cylindrical radial coordinate, and the local
angular velocity $\Omega(\varpi,z,\varphi) \equiv v_{\hat{\varphi}}/\varpi$, where
$v_{\hat{\varphi}}$ is the proper azimuthal component of the fluid
velocity. We also define the average differential rotation $\Delta \Omega$
as the weighted variance of $\Omega$,
\begin{eqnarray}
\bigl( \Delta\Omega \bigr)^2 &=&
I^{-1} \int \rho \varpi^2 \bigl( \Omega - \bar{\Omega} \bigr)^2
d^3 x\nonumber\\
&=&
I^{-1} \int \rho \varpi^2 \Omega^2 d^3 x - \bar{\Omega}^2.
\label{eq:diffrot}
\end{eqnarray}

\section{\label{reaction}Radiation-Reaction Force}

The gravitational radiation-reaction force due to a time-varying current
quadrupole is given by the expression
\begin{eqnarray}
\!\!\!\!\!\!\!\!\!
F^{\mathrm{GR}}_a &=& \kappa \frac{16}{45} \frac{G}{c^7}
\biggl( 2 v_j \epsilon_{jal} x_m S^{(5)}_{lm}
           + v_j \epsilon_{jkl} x_k S^{(5)}_{la} \nonumber \\
        &&\qquad\quad - v_j \epsilon_{akl} x_k S^{(5)}_{lj}
           - \epsilon_{akl} x_k x_m S^{(6)}_{lm} \biggr),
\label{eq:rrforce}
\end{eqnarray}
see Blanchet~\cite{Blanchet97}, and Eq.~(20) of Rezzolla \emph{et al.}\ 
\cite{Rezzolla99}. Here $S^{(n)}_{jk}$ represents the
$n$'th time derivative of the current quadrupole tensor,
\begin{equation}
S_{jk}=\int \rho(\vec{x}\times\vec{v})_{(j}x_{k)} d^3 x;
\label{eq:curquad}
\end{equation}
$\epsilon_{jkl}$ is the totally antisymmetric tensor, and the
vector $x_k$ represents the Cartesian coordinates of the point at
which the force is evaluated. The parameter $\kappa$ that appears
in Eq.~\eqref{eq:rrforce} has the value $\kappa=1$ in general
relativity.  For reasons discussed below, we find it useful to
consider other values of $\kappa$ in our numerical simulations.

We find that a straightforward application of
Eq.~\eqref{eq:rrforce} in numerical evolutions is nearly
impossible.  There are two problems: first, it is very hard to
evaluate reliably time derivatives of such a high order; second,
various sources of numerical noise (even small errors in the
initial equilibrium configuration of the fluid, and the numerical
drift of the center of mass) can generate contributions to the
current quadrupole tensor that overwhelm those of the pure
$r$-mode motion. So we need to introduce special numerical
techniques and simplifications to overcome these problems.

In order to reduce the influence of extraneous noise sources on
the evolution, it is helpful to re-express the current quadrupole
tensor in terms of the current multipole moments defined in
Eq.~(\ref{eq:j22}). There is a one-to-one correspondence between
the $J_{2m}$ current multipoles and $S_{ij}$:
\begin{eqnarray}
S_{yy} - S_{xx} + 2 i S_{xy} & = & \sqrt{\frac{16 \pi}{5}} J_{22}, \label{eq:sandja} \\
S_{xz} - i S_{yz} & = & \sqrt{\frac{4\pi}{5}} J_{21}, \label{eq:sandjb} \\
S_{xx} + S_{yy} = - S_{zz} & = & \sqrt{\frac{8\pi}{15}}J_{20}. \label{eq:sandjc}
\end{eqnarray}
In a slowly rotating star, the $m=2$ $r$-mode excites $J_{22}$,
but not $J_{21}$ and $J_{20}$. In contrast, the principal sources
of numerical noise contribute primarily to $J_{20}$.  Thus, we
evaluate only the $J_{22}$ contribution to
$\vec{F}^{\mathrm{GR}}$: we use Eq.~(\ref{eq:rrforce}) to evaluate
$\vec{F}^{\mathrm{GR}}$, taking the $S_{ij}$ determined from
Eqs.~\eqref{eq:sandja}--\eqref{eq:sandjc}, but setting
$J_{21}=J_{20}=0$.  We find that this scheme reduces considerably
the numerical noise in the radiation reaction force, and
reproduces faithfully the analytical description of $r$-modes in
slowly rotating stars (see Sec.~\ref{calibration}).

The second major problem is evaluating the numerical time
derivatives of $S_{jk}$, or equivalently the time derivatives of
$J_{22}$. Whenever the radiation-reaction timescale is much longer
than the $r$-mode period $2 \pi/\omega$, the dominant contribution
to the derivatives $S^{(n)}_{jk}$ comes from terms proportional to
powers of the $r$-mode frequency:
\begin{equation}
S^{(n)}_{jk} \approx (i\omega)^n S_{jk}, \quad J^{(n)}_{22}
\approx (i\omega)^n J_{22}. \label{eq:sincase}
\end{equation}
Even when the $r$-mode amplitude becomes large, the expression
\eqref{eq:sincase} will be accurate as long as the timescale for
the evolution of $\alpha$ and $\omega$ is longer than $2
\pi/\omega$. Now, $J_{22}$ and $J^{(1)}_{22}$ are easily evaluated
using the integral expressions in Eqs.~(\ref{eq:j22}) and
(\ref{eq:dj22}); thus, the time derivatives needed in
Eq.~(\ref{eq:rrforce}) are given simply by $J^{(5)}_{22}=\omega^4
J^{(1)}_{22}$ and $J^{(6)}_{22}=-\omega^6 J_{22}$, where we
determine $\omega$ numerically using Eq.~\eqref{eq:genomega}.  In
the Appendix, we present explicit expressions for the
components of the effective radiation-reaction force in
cylindrical coordinates.

\section{\label{calibration}Calibration runs}

In order to test the accuracy of our hydrodynamic evolution code
and of our approximations for the gravitational radiation-reaction
force, we investigate the evolution of a small-amplitude $r$-mode
in a slowly rotating star.

We provide initial data for this study by solving the
time-independent fluid equations for a slowly, rigidly rotating
stellar model.  We model the neutron star as an $n = 1$ polytrope,
generated by the self consistent field technique developed by
Hachisu \cite{Hachisu86}.  Table \ref{table:models} shows the
physical parameters for this model, labeled \emph{Slow}; in particular,
the ratio of rotational kinetic energy to gravitational binding
energy is $T_{\mathrm{rot}}/|W| = 0.00398$, and the angular
velocity is 26\% of the maximum possible value (estimated as
$\Omega_{\mathrm{max}} = \frac{2}{3} \sqrt{\pi G \bar{\rho}}$).
\begin{table}
\caption{\label{table:models}Physical parameters for the equilibrium models.}
\begin{ruledtabular}
\begin{tabular}{lcrr}
Parameter        & Symbol   & \emph{Slow} & \emph{Fast}\\
                 &          & C1, C2      &  C3--C8 \\
\hline
polytropic index      & $n$      & 1        & 1 \\
%poly.\ const.     & $K$      & 0.623953 & 0.296814 \\
total mass        & $M$      & $1.4 \, M_\odot$ &
                               $1.4 \, M_\odot$ \\
equatorial radius    & $R_{\mathrm{eq}}$ & 12.7 km & 18.4 km \\
polar R/equatorial R    & $R_{\mathrm{pol}}/R_{\mathrm{eq}}$
                            & 0.98  & 0.59 \\
nonrotating\footnote{$R_0$ is the radius of the nonrotating star of
the same mass.  For $n=1$ polytropes $R_0=\sqrt{\pi K/2 G}$ \protect\cite{Chandrasekhar67} where $K$ is the polytropic constant.} R
  & $R_0$    & 12.5 km &
                               12.5 km \\
angular velocity     & $\Omega_0$ & 1.45 krad/s &
                              5.34 krad/s \\
rotation period           & $P_0$    & 4.32 ms &
                              1.18 ms \\
energy ratio & $T_{\mathrm{rot}}/|W|$  & $3.98 \times 10^{-3}$  & 0.10072 \\
simulation RR timescale & $\tau^{(s)}_{\mathrm{RR}}$ & $0.459 \, P_0$ & $9.4374 \, P_0$ \\
physical RR timescale & $\tau^{(p)}_{\mathrm{RR}}$ & $2.8 \times 10^7 P_0$ & $4.2 \times 10^4 P_0$ \\
\end{tabular}
\end{ruledtabular}
\end{table}

We then adjust the velocity field of this equilibrium model by
adding the velocity perturbation of an $m=2$ $r$-mode
of amplitude $\alpha_0$:
\begin{equation}
\vec{v} = \varpi\Omega_0\vec{e}_{\hat{\varphi}} + \alpha_0 R\Omega_0
\left(\frac{r}{R}\right)^2 {\mathrm{Re}}(\vec{Y}^{B}_{22}).
\label{eq:initialmodel}
\end{equation}
In the Appendix, we write out explicitly the components
of this initial velocity field in our cylindrical coordinate
system. Because Eq.~\eqref{eq:initialmodel} is the exact
representation of a pure $m=2$ $r$-mode only in the small-amplitude,
small-rotation limit, we expect that the frequency and the amplitude
measured in our numerical experiment using
Eqs.~\eqref{eq:alphadef}--\eqref{eq:genomega} will be different from
their theoretical values $\alpha_0$ and $-\frac{4}{3} \Omega_0$ by terms
of order $O(\alpha^2)$, and $O(\Omega^2/\Omega_{\mathrm{max}}^2)$.
%%Following line NOT NECESSARY because defined in previous paragraph.
%%[where $\Omega_{\mathrm{max}} = 2\sqrt{\pi G \bar{\rho}}/3$].

We perform two numerical integrations of the equations of motion, for
this slowly rotating initial configuration. In the first run (C1), we
let the star evolve under purely Newtonian hydrodynamics, setting the
strength $\kappa$ of the radiation reaction \eqref{eq:rrforce} to
zero. In the second run (C2), we force the mode by setting $\kappa
\simeq 6 \times 10^7$.  With this unphysically large value the
amplitude of the $r$-mode grows appreciably within a time that we can
conveniently follow numerically.  (The Courant limit for the evolution
timestep is set by the speed of sound in the fluid, and by the size of
the grid cells; for $\Omega_0 = 0.26 \Omega_{\mathrm{max}}$, one complete
rotation of the star takes about 70000 timesteps).
\begin{figure}
\begin{center}
\includegraphics{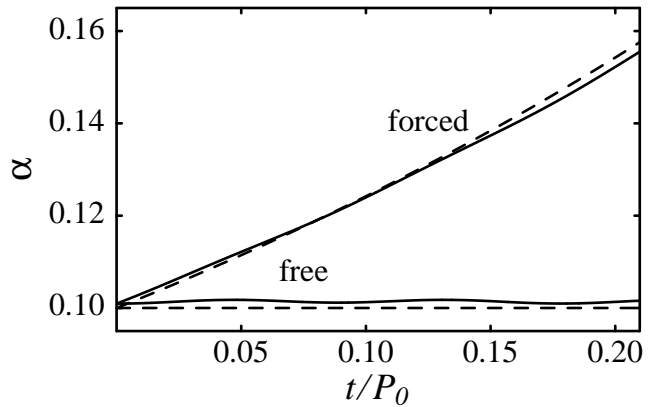} \caption{Evolution of the
$r$-mode amplitude in a slowly rotating star.  The solid curves
plot the results of numerical evolutions (with and without
gravitational radiation reaction) while the dashed curves plot the
analytical predictions.  For the curves marked ``free,'' $\kappa =
0$; for the curves marked ``forced,'' $\kappa \simeq 6 \times
10^7$. \label{fig:slowfreeforcedamp}}
\end{center}
\end{figure}

Figure~\ref{fig:slowfreeforcedamp} illustrates the evolution of
the mode amplitude $\alpha$ in runs C1 and C2, as a function of
$t/P_0$, where $P_0=2\pi/\Omega_0$ is the initial rotation period of
the star. The solid curves trace the numerical evolution of
$\alpha$ [as defined in Eq.~\eqref{eq:alphadef}], whereas the
dashed curves trace the theoretical predictions for this
evolution, obtained in the small-amplitude, slow-rotation limit
\cite{Lindblom98}.

When $\kappa=0$, the theoretical prediction for the evolution of
the amplitude is just $\alpha=\alpha_0$, and we can verify that
the numerical evolution tracks the analytical curve within the
expected deviations of order $\alpha^2$. When $\kappa \neq 0$, the
analytical prediction for the evolution of the amplitude is
\begin{equation}
\alpha = \alpha_0 e^{t/\tau_{GR}},
\label{eq:thevol}
\end{equation}
where the radiation-reaction timescale is given by \cite{Lindblom98}
\begin{equation}
\frac{1}{\tau_{\mathrm{GR}}} = 2\pi \left(\frac{256}{405}\right)^2
\kappa \frac{G}{c^7}
\tilde{J} M R^4 \Omega_0^6.
\label{eq:rrtimescale}
\end{equation}
For this model, $\tau_{\mathrm{GR}} = 0.46 P_0$. As we can see in
Fig.~\ref{fig:slowfreeforcedamp}, the numerical evolution tracks
the small-amplitude, slow-rotation analytical result within the
expected accuracy, even if the radiation-reaction force is so
unphysically strong.

Although this slow rotation numerical evolution was only carried
out over a small fraction (0.2) of a rotation period (and
therefore over a small fraction of the $r$-mode oscillation
period), the evolution extended for about 7.3 dynamical times and
4.6 sound-crossing times.
\begin{figure}
\begin{center}
\includegraphics{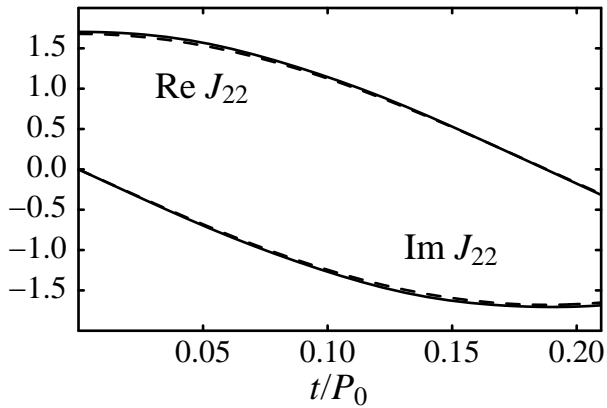} \caption{Real and imaginary parts of
the current multipole moment $J_{22}$ (in arbitrary units), for a
slowly rotating star evolved without gravitational radiation
reaction (run C1).  The solid curves trace the numerical evolutions,
the dashed curves trace the analytical predictions.\label{fig:slowfreej}}
\end{center}
\end{figure}

\begin{figure}
\begin{center}
\includegraphics{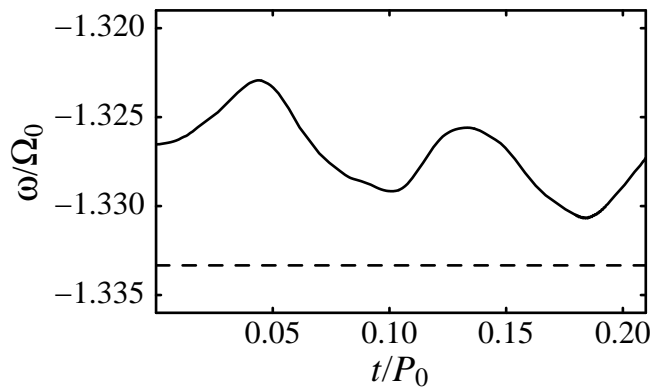} \caption{Frequency of the $m=2$
$r$-mode, for a slowly rotating star evolved without gravitational
radiation reaction (run C1).  The solid curve is determined
numerically from Eq.~(\ref{eq:genomega}). For comparison, the
dashed line shows the analytical value
$\omega_0=-\frac{4}{3}\Omega_0$. \label{fig:slowfreeomg}}
\end{center}
\end{figure}

In Figs.~\ref{fig:slowfreej} and \ref{fig:slowfreeomg}, we display
two additional diagnostics for the undriven ($\kappa=0$)
slow-rotation evolution (C1).  In Fig.~\ref{fig:slowfreej} we plot
the real and imaginary parts of the current multipole moment
$J_{22}$: the solid curves trace the numerical evolution, whereas
the dashed curves trace the analytical expression
\begin{equation}
J_{22}=\frac{1}{2}\alpha M R^3 \tilde{J} \Omega_0 e^{i\omega t}.
\end{equation}
Again the deviations are within the expected accuracy of the
analytical results.  The deviations appear to be caused by the
excitation of modes other than the pure $m=2$ $r$-mode; the spurious
excitations appear because the initial data
[Eq.~\eqref{eq:initialmodel}] are only accurate to first order in
$\alpha$.  Figure~\ref{fig:slowfreeomg} depicts the evolution of the
frequency $\omega$ [as defined in Eq.~\eqref{eq:genomega}].  The
deviations from the analytical result, $\omega_0=-\frac{4}{3}\Omega_0$,
are within the expected accuracy. The magnified scale used to display
$\omega$ in Fig.~\ref{fig:slowfreeomg} makes the presence of the
small-amplitude, short-period extraneous modes quite apparent.

\section{\label{sec:fast}Evaluating the saturation amplitude}

In our production runs, we investigate the nonlinear behavior of
the $r$-mode in a rapidly rotating stellar model, under a variety
of physical conditions (different initial amplitudes, and
different values for the radiation-reaction coefficient $\kappa$).

Again, we provide initial data by solving the time-independent
fluid equations for an $n = 1$ polytrope. The physical parameters
for this model, labeled \emph{Fast}, are reported in Table
\ref{table:models}; in particular, the ratio of rotational kinetic
energy to gravitational binding energy is $T_{\mathrm{rot}}/|W| =
0.10072$, and the angular velocity is 95\% of its maximum value.

We perform a numerical integration of the equations of motion
starting from the rapidly rotating initial configuration,
\emph{Fast}, using Eq.~\eqref{eq:initialmodel} to add a
slow-rotation, small-amplitude $r$-mode field, with $\alpha_0 =
0.1$. Because the radiation-reaction force is so much stronger for
this model (it is proportional to $\omega^6 \propto \Omega^6$), we
find that we can set $\kappa = 4487$, which yields an $r$-mode
growth time $\tau^{(s)}_{\mathrm{RR}} = 9.43 P_0$. This choice of
$\kappa$ is still much larger than its physical value (unity), but
it should yield a reasonable picture of the nonlinear evolution of
the $r$-mode, if the timescales for all the relevant
hydrodynamical processes (including nonlinear couplings to other
modes) are comparable to $P_0$, or shorter. Indeed, if the average
sound-crossing time $\tau_S$ is representative of the relevant
hydrodynamical timescales, then our condition is satisfied: a
rough estimate gives $\tau_S = R_0 / \bar{c}_S \simeq 0.16 P_0
\simeq \tau^{(s)}_{\mathrm{RR}}/60$, where we have approximated $\bar{c}_S$ as the average speed of sound in the equivalent spherical polytrope.
\begin{figure}
\begin{center}
\includegraphics{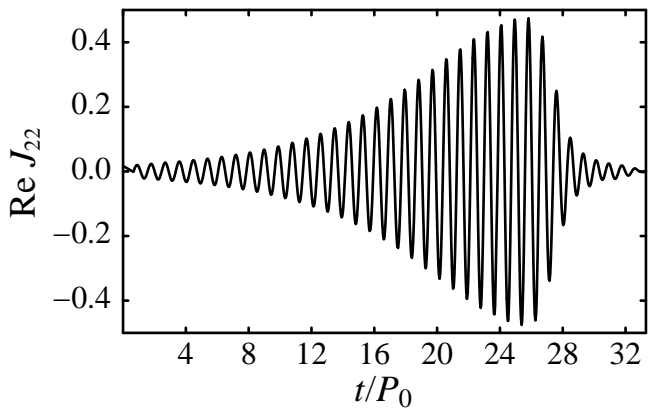} \caption{Numerical evolution of
$\mathrm{Re} \, J_{22}$ (arbitrary units) for a rapidly rotating
star driven by gravitational radiation reaction (production run
C3). The sinusoidal approximation, used to compute $\omega$ and
$J_{22}^{(n)}$, is evidently appropriate for this run.
\label{fig:fastj}}
\end{center}
\end{figure}
\begin{figure}
\begin{center}
\includegraphics{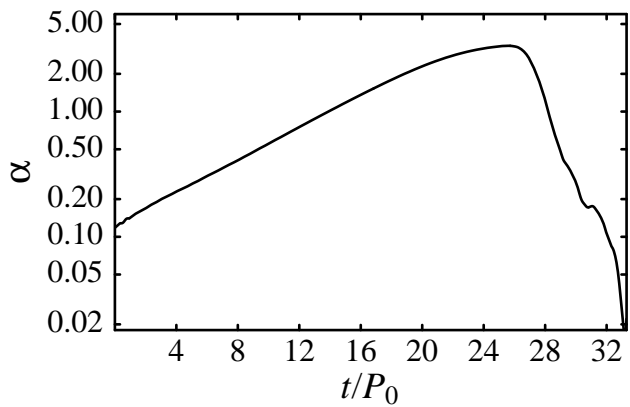} \caption{Numerical evolution of the
$r$-mode amplitude $\alpha$ in production run C3.
\label{fig:fastamp}}
\end{center}
\end{figure}

\subsection{Evolution of the $r$-mode amplitude}

We follow the evolution through $t = 33 P_0$. Because the rotation
of the star is progressively reduced by radiation reaction, and
because the star develops differential rotation, at the end of the
evolution the star has performed, \emph{on the average}, only
about 31 rotations [we obtain this number from $\int \bar{\Omega}
dt / (2 \pi)$]. In Fig.~\ref{fig:fastj} we plot the numerically
determined evolution of ${\rm Re} \, J_{22}$: the curve is a very
smooth sinusoid, whose frequency is essentially constant, and
whose envelope is determined by the (relatively slow) evolution of
the $r$-mode amplitude. So the approximations used to compute
$\omega$ and $J_{22}^{(n)}$ (discussed in Sec.~\ref{reaction}) are
in fact quite good in this situation.

In Fig.~\ref{fig:fastamp} we plot the numerical evolution of the
$r$-mode amplitude $\alpha$. At the beginning of the evolution,
the computed diagnostic $\alpha$ agrees with the theoretical value
$\alpha_0$ to $\sim$ 10\%, within the expected accuracy. The
growth is exponential (as predicted by perturbation theory) until
$\alpha \approx 1.8$. Then some nonlinear process begins to limit the
growth, until the amplitude peaks at $\alpha=3.35$ and then falls
rapidly within a few rotation periods.  After this the $r$-mode is
effectively not excited.

\subsection{\label{subsec:saturation}A mechanism for $r$-mode saturation}

\begin{figure}
\begin{center}
\includegraphics{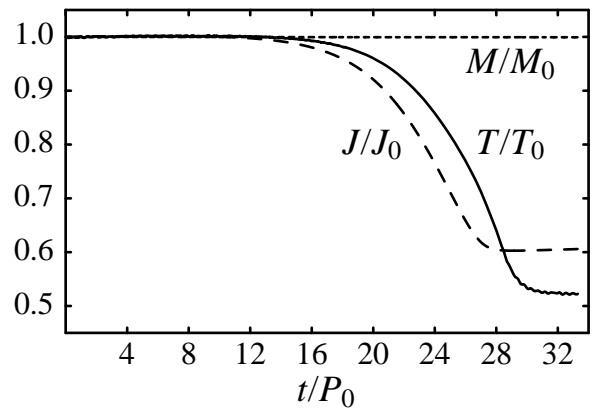} \caption{Evolution of total mass,
total angular momentum, and total kinetic energy in production run
C3. The quantities are plotted as fractions of their initial
value. \label{fig:fastintegrals}}
\end{center}
\end{figure}

What nonlinear process is responsible for the behavior of the $r$-mode
amplitude?  What causes the mode to saturate and disappear from
the star?  To answer these questions, we study the evolution of the
total mass, total angular momentum, and total kinetic energy of the
star, which are plotted in Fig.~\ref{fig:fastintegrals}.

\begin{figure}
\begin{center}
\includegraphics{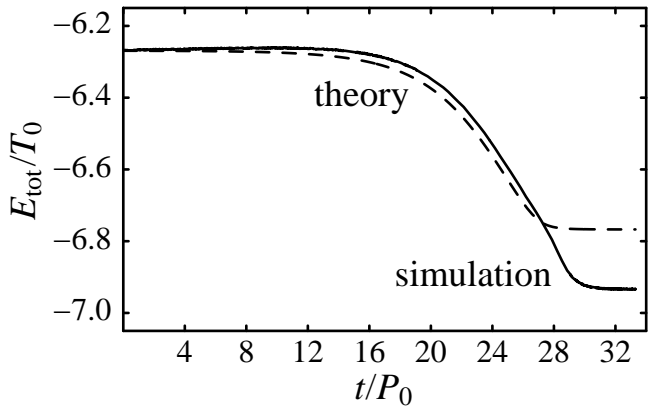} \caption{Theoretical and numerical
evolution of the total energy for production run C3. The total
energy is plotted in units of the initial rotational energy
(because the system is bound, $E_{\mathrm{tot}}$ must be
negative). \label{fig:fastenergy}}
\end{center}
\end{figure}

Because the mass is constant, the damping of the $r$-mode cannot be
caused by ejection of matter from the simulation grid. On the other
hand, we expect that the star should lose energy and angular momentum
as it radiates gravitational radiation in accord with the prediction
of general relativity \cite{Rezzolla99,Thorne80}:
\begin{equation}
\left( \frac{dE}{dt} \right)_{\! J_{22}} =
\frac{|\omega|}{2} \left( \frac{dJ}{dt} \right)_{\! J_{22}}
= - \frac{128\pi}{225} \frac{G}{c^7} \kappa \omega^6 |J_{22}|^2.
\label{eq:energyloss}
\end{equation}
The evolution of the angular momentum mirrors this equation quite
closely (within a few percent); for energy, however,
Eq.~\eqref{eq:energyloss} is only accurate until shortly after the
catastrophic fall of the $r$-mode amplitude (at $t \simeq 28 P_0$; see
Fig.~\ref{fig:fastenergy}). Before that time, the star loses about
40\% of its initial angular momentum and 36\% of its initial kinetic
energy. After that time, the amplitude and (therefore) the
radiation-reaction force are much reduced, so $J$ becomes essentially
constant; however, the kinetic energy continues to decrease, losing an
additional 12\% of its initial value during the next three rotation
periods.

If the $r$-mode were damped by a hydrodynamical process that
conserved energy, such as the transfer of energy to other modes,
then Eq.~\eqref{eq:energyloss} should portray accurately the
evolution of the kinetic energy. But this is not what we see:
instead, some \emph{purely hydrodynamic} process continues to
decrease the energy (by a sizable amount!) after the
gravitational-radiation losses become negligible.

\begin{figure}
\begin{center}
\includegraphics[width=3.4in]{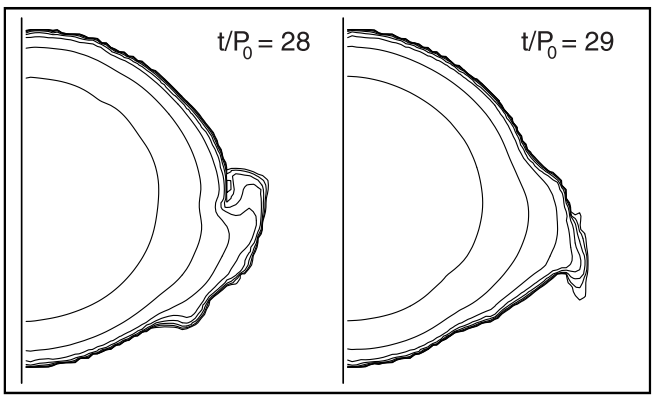}
\caption{Isodensity surfaces showing breaking waves near the end of
production run C3.\label{fig:fastwaves}}
\end{center}
\end{figure}

We believe that we have identified this process. To first order in the
amplitude, the $r$-mode is only a \emph{velocity} mode; to second
order, however, there is also an associated density perturbation,
proportional to $Y_{32}$, which appears as a wave with four crests
(two in each hemisphere) on the surface of the star. (We will present
a quantitative analysis later in this section.) As the amplitude
reaches its maximum, these crests become large, breaking waves: the
edges of the waves develop strong shocks that dump kinetic energy into
thermal energy.  In doing so they damp the
$r$-mode. Figure~\ref{fig:fastwaves} illustrates the surface waves at
$t = 28 P_0$ and $t = 29 P_0$ along selected meridional slices.

Our code is written in such a way that the evolution of the shocks
is always \emph{kinematically} correct. In particular, the
continuity equation ensures that mass is properly conserved, and
that the proper density jump occurs across the shocks; and the
Euler equation ensures that momentum is properly conserved, and
that the proper pre- and post-shock velocities arise. However, our
code does not allow the energy dissipated in the shock to increase
the entropy of the fluid, which remains always barotropic and
isentropic. Consequently, the presence of shocks shows up as a
decrease in the total energy of the star. Indeed, in this
production run (C3) gravitational radiation reduces the
\emph{total} energy by 9\% of its initial magnitude through $t =
28 P_0$, and dissipation in shocks subtracts a further 3\% in the
last three rotations. (For comparison, 12\% of the initial kinetic
energy is lost in the last three rotations compared to 36\% before
that time.)

Ignoring the thermal effects of shocks is useful to reduce the
computational burden and the complexity of the hydrodynamic code,
and it is in fact a fairly reasonable approximation for neutron
star matter, where the pressure comes mostly from the Fermi
pressure of the degenerate neutrons, so the equation of state can
be effectively modeled as temperature-independent.  

\subsection{Radial structure of the $r$-mode amplitude}

We define the radial amplitude density $\alpha(r)$ (where $r$ is
the \emph{spherical} radius) by expressing the integral
Eq.~\eqref{eq:j22} for $J_{22}$ in spherical coordinates, and
omitting the radial integration:
\begin{equation}
\alpha(r) e^{i \phi(r)} = \frac{2}{\tilde{J}MR^2\Omega_0}
\int \rho r^2 \vec{v}\cdot\vec{Y}^{B*}_{22} r^2 \sin\theta d\theta \, 
d \varphi.
\end{equation}
We removed the absolute value around the integral for $J_{22}$ so
that we can keep track of the local mode phase $\phi(r)$. With
this definition, $\alpha \exp [i \phi] =  \int \alpha(r) \exp [i
\phi(r)] dr/R$, where $\phi$ is the global phase of the $r$-mode.

\begin{figure}
\begin{center}
\includegraphics{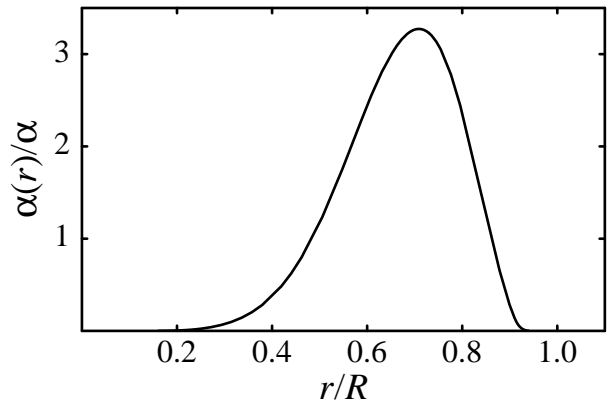} \caption{Radial amplitude density
$\alpha(r)$ of the $m=2$ $r$-mode for production run C3 at $t =
22.5P_0$. \label{fig:rmodestructure}}
\end{center}
\end{figure}

The amplitude $\alpha(r)$ is plotted in Fig.~\ref{fig:rmodestructure}
for the production run C3 at the time $t = 22.5P_0$.  Throughout the
entire evolution, the mode is concentrated mostly between the
spherical radii $r = 0.5 R$ and $0.9 R$, and the shape of $\alpha(r)$
is fitted reasonably well by taking $\delta v \propto (r/R)^2$ [see
Eq.~\eqref{eq:rmodedef}] and $\rho \propto (\sin \pi r/R) / (\pi r/R)$
as appropriate for a spherical, $n=1$ polytrope.

\begin{figure}
\begin{center}
\includegraphics{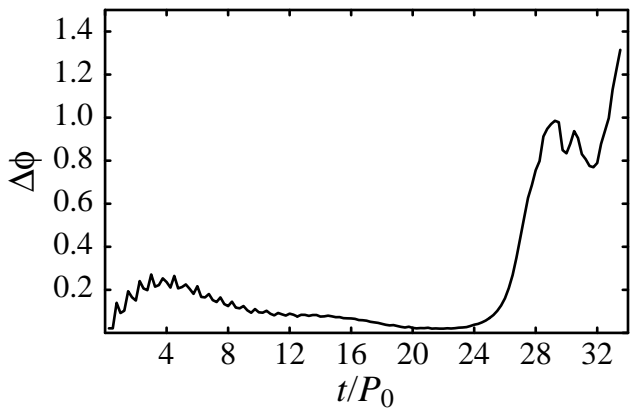} \caption{Evolution of the
phase-coherence function $\Delta \phi$ in production run C3.
\label{fig:rmodecoherence}}
\end{center}
\end{figure}

It is also interesting to study the phase coherence of the
$r$-mode, which we define as
\begin{equation}
(\Delta \phi)^2 =
\frac{
\int \alpha(r)
\bigl| e^{i \phi(r)} - e^{i \phi} \bigr|^2 dr}{
\int \alpha(r) dr}.
\end{equation}
Figure \ref{fig:rmodecoherence} plots the evolution of $\Delta
\phi$, which is small until the $r$-mode saturates at $t \approx 26
P_0$. For $\Delta \phi \approx 1$, the local phase $\phi(r)$ spans
approximately $2 \pi$: the mode has lost coherence completely. In
this situation, there are large regions in the star where the
radiation-reaction force pushes out of phase with the local mode
oscillations; this mismatch accelerates the damping of the mode.

\subsection{Evolution of the $r$-mode frequency}

\begin{figure}
\begin{center}
\includegraphics{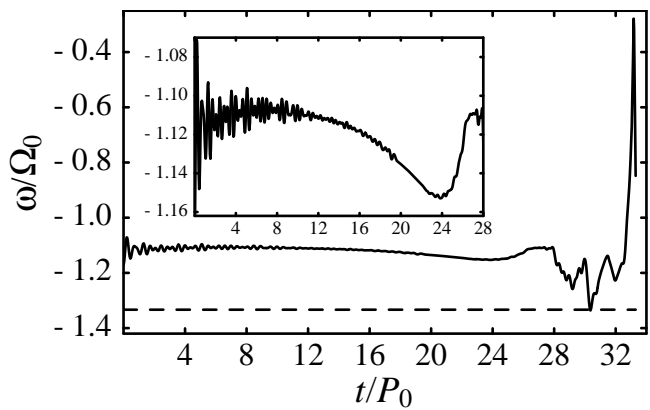} \caption{Numerical evolution of the
$r$-mode frequency $\omega$ in production run C3.
\label{fig:fastomg}}
\end{center}
\end{figure}

Figure \ref{fig:fastomg} shows the numerical evolution of the
$r$-mode frequency $\omega$.  The evolution of $\omega$ is quite
smooth when the amplitude of the $r$-mode is large; when the
amplitude is small (for $t \lesssim 10 P_0$ and for $t \gtrsim 28
P_0$), we see that other modes make noticeable contributions to
$J_{22}$, and therefore to $\omega$. At the beginning of the run,
the numerical $\omega$ matches the theoretical prediction to within
the expected accuracy of about 10\%.  These values for the frequency
are also consistent with those obtained via a Fourier transform of
$J_{22}$~\cite{owen_lindblom01}. 

\begin{figure}
\begin{center}
\includegraphics{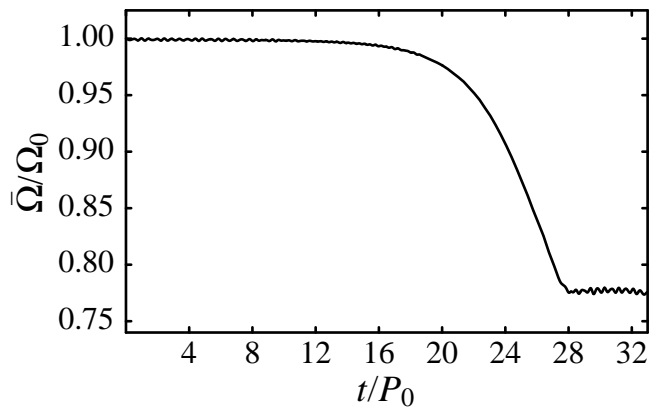} \caption{Numerical evolution of the
average stellar angular velocity $\bar{\Omega}$ in production run
C3. \label{fig:fastbaromg}}
\end{center}
\end{figure}

Surprisingly, the $r$-mode frequency remains approximately
constant throughout the evolution, and it does not follow the
decline of the average angular velocity $\bar{\Omega}$ (plotted in
Fig.~\ref{fig:fastbaromg}). Altogether, the angular velocity
decreases by about 22.5\% while the total angular momentum decreases by
40\%.  (As the star spins down, it becomes less flattened,
and the change in the moment of inertia  accounts for the
difference between the decrease of $J$ and that of
$\bar{\Omega}$.)  The stability of the $r$-mode frequency has
important implications for the possible detection of $r$-mode
gravity waves (see Sec.~\ref{conclusion}).

We also point out that the approximate expressions for the GR reaction
force, Eqs.~\eqref{eq:rrforce}--\eqref{eq:sincase}, that we use here
are accurate only when the motion of the fluid has nearly sinusoidal
time dependence.  Figure~\ref{fig:fastomg} illustrates that the
evolution in our simulation remains quite sinusoidal until about
$t=28P_0$.  After this point our expression for the GR reaction force
is not reliable.  After this point in our simulation, however, the
fluid evolution is dominated by nonlinear hydrodynamic forces
including shocks, and the GR reaction force is negligible.  Thus our
inability to model accurately the GR force during the late stages of
the evolution does not effect our results.

\subsection{Growth of differential rotation}

\begin{figure}
\begin{center}
\includegraphics{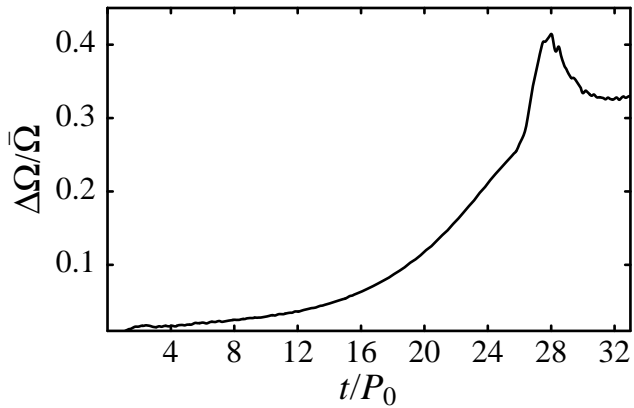} \caption{Numerical evolution of the
differential rotation $\Delta \Omega$ in production run C3.
\label{fig:fastdelomg}}
\end{center}
\end{figure}

During this simulation (run C3), the average differential rotation
$\Delta\Omega$ [defined in Eq.~\eqref{eq:diffrot}] grows to a maximum
of approximately $0.41 \bar{\Omega}$ (see
Fig.~\ref{fig:fastdelomg}). After a rapid increase in the first three
rotation periods, when the linear $r$-mode eigenfunction of
Eq.~\eqref{eq:initialmodel} evolves into its proper nonlinear,
rapid-rotation form, $\Delta\Omega/\bar{\Omega}$ increases
approximately as $\alpha^{0.75}$ until $\alpha \simeq 1$, and then
approximately as $\alpha$ until $\alpha$ begins to saturate. When
$\alpha$ is maximum, $\Delta\Omega = 0.25 \bar{\Omega}$. As the
amplitude falls, $\Delta\Omega$ continues to grow (even more steeply),
as long as there is significant gravitational radiation. After $t = 28
P_0$, $\Delta\Omega$ decreases to about 80\% of its peak value. So the
final configuration of the star (where the presence of the $r$-mode is
essentially negligible) still has a very large differential rotation.
\begin{figure}
\begin{center}
\includegraphics[width=3.4in]{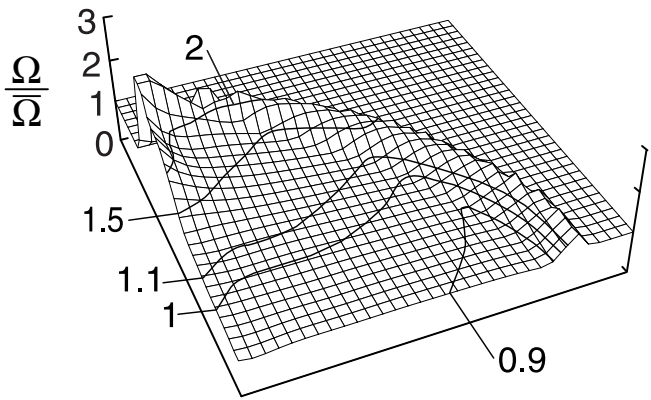} \caption{Meridional
structure of the differential rotation in production run C3. The plot
shows the value of the azimuthally averaged angular velocity
$\Omega(\varpi,z)/\bar{\Omega}$, at time $t = 25.6 P_0$.
\label{fig:fastdelstruct}}
\end{center}
\end{figure}

But we should not concentrate exclusively on the averaged quantity
$\Delta\Omega/\bar\Omega$, which does not capture fully the
spatial structure of differential rotation.
Figure~\ref{fig:fastdelstruct} illustrates the spatial dependence
of the azimuthally averaged angular velocity,
\begin{equation}
\Omega(\varpi,z) = \frac{1}{2\pi} \int \Omega(\varpi,z,\varphi) \, d\varphi,
\label{eq:omegavarpi}
\end{equation}
at the time when the amplitude is maximum, $t = 25.6 P_0$.  The
differential rotation is confined mostly to a thin shell of
material near the surface of the star, and is particularly
concentrated near each polar cap. The bulk of the material in the
star remains fairly rigidly rotating.

\subsection{Consistency of the radiation-reaction force}

In these simulations we have assumed that the only relevant
contribution to the radiation reaction force comes from the current
quadrupole moment, and in particular from $J_{22}$. However, in the
post-Newtonian approximation to general relativity, the lowest-order
contribution to radiation reaction comes from the mass quadrupole
term, followed by mass octupole and current quadrupole. To verify that
our approximation is justified for the physical states considered here,
we evaluate the additional energy that would have been lost to
gravitational waves throughout our simulation if we had included the
lowest-order mass multipole terms.

The mass multipole moments $Q_{lm}$ are defined by
\begin{equation}
Q_{lm} = \int \rho r^l Y^*_{lm} d^3x.
\end{equation}
In the presence of density oscillations with sinusoidal
dependences in the coordinates $t$ and $\varphi$ ({\it i.e.},
$\delta \rho_{lm} \propto e^{i \omega_{lm} t + i m \varphi}$) the flux
of energy into gravitational waves is given by
\cite{Thorne69,Ipser91}
\begin{eqnarray}
\label{eq:energylossq2}
\left( \frac{dE}{dt} \right)_{Q_{2m}} \!\!\!\!\! & = & -\frac{8 \pi}{75}
\frac{G}{c^5} \omega_{2m}^6 |Q_{2m}|^2; \\
\left( \frac{dE}{dt} \right)_{Q_{3m}} \!\!\!\!\! & = & -\frac{8 \pi}{6615}
\frac{G}{c^7} \omega_{3m}^8 | Q_{3m}|^2.
\label{eq:energylossq3}
\end{eqnarray}
where $Q_{2m}$ and $Q_{3m}$ are, respectively, the
mass quadrupole and mass octupole moments induced by these density
fluctuations.  Contributions of higher order are suppressed by very
small fractional coefficients.

Comparing Eqs.~\eqref{eq:energylossq2} and \eqref{eq:energylossq3} with
Eq.~\eqref{eq:energyloss} we find that the contribution of the density
oscillations associated with the $r$-mode at frequency $\omega$ to the
energy flux is negligible whenever
\begin{equation}
\frac{3 c^2}{16} \frac{|Q_{2m}|^2}{|J_{2m}|^2} \ll 1, \quad
\frac{5 \omega^2}{2352} \frac{|Q_{3m}|^2}{|J_{2m}|^2} \ll 1.
\end{equation}
We find that in our simulation both ratios are of order $10^{-3}$
before the $r$-mode saturates (at $t \simeq 25 P_0$). The
strongest contribution to the quadrupole term comes from
$Q_{22}$, although the Fourier transform of this moment does not
show any definite frequency of oscillation. The strongest
contribution to the octupole term comes from the $Y_{32}$ 
dependence of the density in the $m=2$ $r$-mode
(see the next subsection).

Between $t \simeq 25 P_0$ and $t \simeq 32 P_0$ (when $\alpha$ is back
to its initial value $\approx 0.1$) the mass quadrupole term would
have provided a correction of order 10\% to the current quadrupole;
although even then we see no evidence of a definite oscillation
frequency correlated to the $r$-mode. Only after $t \simeq 32 P_0$,
when the fluid motion in the star becomes quite turbulent and the
$r$-mode is very weak, is the gravitational radiation generated by the
mass multipoles comparable to the radiation from $J_{22}$.

On the whole, we find that our approximation which ignores the
contributions from the mass multipoles is well-justified throughout
the more interesting part of the evolution.

\subsection{\label{subsec:densper}Density oscillations and mode saturation}

The evolution of the isodensity surfaces in our neutron star shows very
clearly the presence of the lowest-order Eulerian density perturbation
$\delta \rho$ associated with the $m=2$ $r$-mode.  The lowest order
expression for $\delta \rho$ was derived by Lindblom, Owen and Morsink
\cite{Lindblom98} in the small-amplitude, slow-rotation approximation.
Solving Eq.~(5) of Ref.~\cite{Lindblom98} with $m = 2$ and with
polytropic index $n=1$, and then substituting $\delta \Psi$ back into
Eq.~(4) of Ref.~\cite{Lindblom98}, we get
\begin{equation}
\delta \rho = \alpha_0 \frac{7 \pi^2 }{15} \sqrt{\frac{2}{3}}
\frac{\Omega_0^2}{G}j_3\left({\pi r\over R_0}\right) 
Y_{32}(\theta,\varphi) e^{i \omega t},
\label{eq:rhopert}
\end{equation}
where $j_3$ is the spherical Bessel function.
The mass multipole associated with this $\delta \rho$ is
\begin{equation}
\delta Q_{32} = \alpha_0 \frac{7\pi}{15} \sqrt{\frac{2}{3}}
\frac{\Omega_0^2 R_0^5}{G} j_4(\pi) Y_{32}(\theta,\varphi) e^{i \omega t},
\label{eq:rhopertmoment}
\end{equation}
where $j_4(\pi) = 0.151425$.

\begin{figure}
\begin{center}
\includegraphics{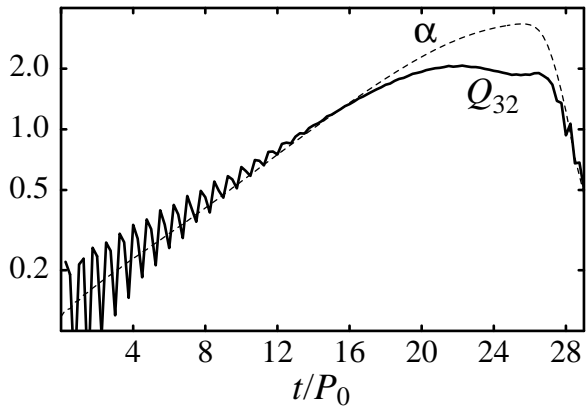} \caption{Numerical evolution of the mass
moment $Q_{32}$ (solid line) and of the $r$-mode amplitude
$\alpha$, in production run C3. The curve for $Q_{32}$ was renormalized to emphasize the linear relation between $\alpha$ and $Q_{32}$ during the growth of the $r$-mode.\label{fig:y32comp}}
\end{center}
\end{figure}

We study the evolution of $Q_{32}$ throughout run C3. We find that
$Q_{32}$ (and therefore the density perturbation with angular
dependence given by $Y_{32}$) is indeed proportional to $\alpha$, at
least as long as the growth of $\alpha$ remains exponential; after
that, $Q_{32}$ grows more slowly than $\alpha$, and it reaches a
maximum a few rotation periods before $\alpha$ (see
Fig. \ref{fig:y32comp}).  The phase evolution of the density
perturbation is also consistent with expectations: the Fourier
transform of $Q_{32}(t)$ shows a very definite peak at the $r$-mode
(numerical) frequency $\omega$.

A quantitative check shows that Eq.~\eqref{eq:rhopertmoment} predicts
the observed magnitude of $Q_{32}$ with an accuracy of about 50\%;
this error is consistent with the next-order terms ($\sim \Omega^4$
and $\alpha^2$) not included in this expression.  In the slowly
rotating calibration model C1, we find that $Q_{32}$ is given by
Eq.~\eqref{eq:rhopertmoment} to within about 1\%.

We point out that we do not explicitly include any density
perturbation in the initial configuration of the star; rather, the
density perturbation is immediately generated by the hydrodynamic
evolution of the fluid as a consequence of the initial velocity
perturbation. The evolution of the amplitude of the density
perturbation amplitude provides more insight into the mechanism that
causes the $r$-mode to saturate: on the surface of the star, $\delta
\rho$ appears as four large wave crests; at a critical amplitude
these crests stop growing, and within a few rotation periods they turn
into breaking waves that damp the $r$-mode.

\subsection{Limits on mode--mode coupling}

In the numerical evolution (C3) nonlinear hydrodynamic processes do
not prevent the gravitational radiation instability from driving the
dimensionless amplitude of the $r$-mode to values of order unity. In
particular, the energy of the $r$-mode is not channeled into other
modes by nonlinear hydrodynamic coupling until the amplitude of the
mode becomes quite large. It is possible however that the nonlinear
processes that would limit the growth of the $r$-mode act only on
timescales that are longer than our artificially brief simulation
growth time $\tau^{(s)}_{\mathrm{RR}}$, but still shorter than the
physical $\tau^{(p)}_{\mathrm{RR}}$. 

Can our numerical simulation place any limits at all on the
possibility of nonlinear coupling?  We know that in our simulation
the amplitude of the $r$-mode grows exponentially until $\alpha
\approx 2$, so the nonlinear interaction with other modes must be
negligible at least until that time.  This observation allows us to
set a limit on the strength of the nonlinear couplings between the
modes; and from this limit we can infer a \emph{lower limit} on the
saturation amplitude that may be achieved when the radiation-reaction
coupling is adjusted to its physical value.  Of course, the inference
is only justified for the nonlinear interaction of the $r$-mode with
other modes that are correctly modeled in our simulation (for
instance, the finite azimuthal resolution of the grid sets an upper
limit on the $m$ of the modes that can be resolved), and with our
physical assumptions (for instance, the buoyant $g$-modes of realistic
neutron stars will not be present with our choice of the equation of
state).

Our argument is based on the Lagrangian description of the nonlinear
evolution of the mode amplitude developed by Schenk, \emph{et al.}\
\cite{Schenk01}.  In this formalism, the modes interact at the lowest
order by way of \emph{three-mode couplings}: roughly speaking,
quadratic interactions between pairs of modes drive the evolution of
the amplitude of a third mode.  Because at the beginning of our
simulation all modes except the $r$-mode have negligible amplitude, we
expect that the most important three-mode nonlinear term might be one
that couples two $r$-modes to a third mode \cite{Schenk01}.  Following
Ref.~\cite{Schenk01} we consider the coupled equations for the
$r$-mode and a generic mode $X$ obtained in second-order Lagrangian
perturbation theory:
\begin{eqnarray}
\frac{d c_R}{dt} + i \omega_R c_R & = & \frac{c_R}{\tau_{\mathrm{RR}}} +
  \frac{i \omega_R}{2} \frac{\kappa^*_{XRR}}{\epsilon_R} c^*_R c^*_X,
  \label{eq:crevol}  \\
\frac{d c_X}{dt} + i \omega_X c_X & = &
  \frac{i \omega_X}{2} \frac{\kappa^*_{XRR}}{\epsilon_X} c^*_R c^*_R,
  \label{eq:cxevol}
\end{eqnarray}
where $c_R$ and $c_X$ are the complex amplitudes (including phases) of
the modes; $\omega_R$ and $\omega_X$ are their frequencies;
$\epsilon_R$ and $\epsilon_X$ are the nonlinear mode energies at unit
amplitude; and $\tau_{\mathrm{RR}}$ is the radiation-reaction
e-folding time of the $r$-mode. Finally, $\kappa^*_{XRR}$ is the
nonlinear interaction energy for unit amplitude modes. Schenk, {\it et
al.} \cite{Schenk01} give expressions for the $\kappa_{XRR}$ of
coupled generic Newtonian modes in rotating stars.  In writing
Eqs.~\eqref{eq:crevol} and \eqref{eq:cxevol} we have omitted the coupling
terms proportional to $\kappa^*_{XXR}$, which are forbidden by a
$z$-parity selection rule \cite{Schenk01}: the $r$-mode has odd
$z$ parity, so it cannot couple quadratically to the mode $X$.

From our numerical evolution C3, we know that the amplitude of
the $r$-mode grows very nearly exponentially until $\alpha \simeq 2$:
\begin{equation}
c_R(t) \simeq c_R(0)
  e^{-i \omega_R t + t / \tau^{(s)}_{\mathrm{RR}}},
\label{eq:simplecrevol}
\end{equation}
where $\tau^{(s)}_{\mathrm{RR}}$ is the artificially short
radiation-reaction timescale used in our simulation. (Although it is
convenient to take $|c_R| \simeq \alpha$, our argument still applies
as long as $|c_R|$ is merely proportional to $\alpha$.) Therefore, we
also know that until $|c_R| \simeq 2$, the second term on the right
side of Eq.~\eqref{eq:crevol} is negligible compared to the first.  In
this case,
\begin{equation}
{1\over \tau^{(s)}_{\mathrm{RR}}}
\gg
  \left| \frac{i \omega_R}{2} \frac{\kappa^*_{XRR}}{\epsilon_R} c^*_X
  \right|.
\label{eq:negligible}
\end{equation}
We now use Eq.~\eqref{eq:simplecrevol} to integrate
Eq.~\eqref{eq:cxevol} and compute $c_X$:
\begin{eqnarray}
\!\!\!\!\!
c_X(t) &=& c_X(0)e^{-i\omega_X t}\nonumber\\
&&\!\!\!\!\!+\frac{i \omega_X}{2} \frac{\kappa^*_{XRR}}{\epsilon_X}
\frac{[c_R^*(t)]^2 - [c^*_R(0)]^2e^{-i\omega_Xt}}
{2 i \omega_R + i \omega_X + 2 / \tau^{(s)}_{\mathrm{RR}}}.
\end{eqnarray}
Now we set $c_X(0)\simeq 0$ and $|c_R(t)| \gg |c_R(0)|$ for the time
late in the simulation when $c_R\simeq 2$, and find
\begin{equation}
|c_X(t)| \simeq \left| \frac{\kappa^*_{XRR}}{\epsilon_X} \right|
\frac{|\omega_X| \tau^{(s)}_{\mathrm{RR}} |c_R^*(t)|^2}{2 \sqrt{
\bigl( \tau^{(s)}_{\mathrm{RR}} \delta \omega \bigr)^2 + 4}},
\label{eq:cx}
\end{equation}
where $\delta \omega \equiv 2 \omega_R + \omega_X$. We define the \emph{resonance index}
$\gamma^{(s)}=|\omega_R/\omega_X| [(\tau_{\mathrm{RR}}^{(s)}\delta\omega)^2+4]^{1/2}$, whose value is close to unity, $\gamma^{(s)}\simeq 1$, when the system is near resonance, $\delta\omega\simeq 0$.
We use this bound on $|c_X(t)|$ in Eq.~\eqref{eq:negligible} to obtain
\begin{equation}
\frac{1}{\tau^{(s)}_{\mathrm{RR}}} \gg
\frac{ |\kappa^*_{XRR} |^2 }{ 4\epsilon_X \epsilon_R }
\omega_R^2 \tau^{(s)}_{\mathrm{RR}}
\frac{|c_R^*(t)|^2}{{\gamma^{(s)}}}. 
\end{equation}
We can rewrite this inequality in terms of the $r$-mode period $P_R =
2 \pi / \omega_R$:
\begin{equation}
\left[ \frac{P_R}{\tau^{(s)}_{\mathrm{RR}}} \right]^2 \gg
\pi^2 \frac{|c_R(t)|^2}{{\gamma^{(s)}}} \left| \frac{ \kappa^*_{XRR} }{ \epsilon_X } \right|^2 \frac{\epsilon_X}{\epsilon_R}.
\label{eq:inequalityb}
\end{equation}
We now set $|c_R(t)|=2$ (the value at which the evolution of the
amplitude begins to show deviation from exponential) and $P_R /
\tau^{(s)}_{\mathrm{RR}}=1/10$ (the value for our simulation), and
obtain
\begin{equation}
\left| \frac{ \kappa^*_{XRR} }{ \epsilon_X } \right|^2 
\frac{\epsilon_X}{\epsilon_R}\ll
\frac{\gamma^{(s)}}{400 \pi^2} .
\label{eq:klimit}
\end{equation}
Thus, our numerical evolution puts a limit on the strength of the
coupling between the $r$-mode and other modes in the star.

We now ask how the saturation amplitude would change if the
radiation-reaction timescale assumed its physical value
$\tau^{(p)}_{\mathrm{RR}}$ instead of the value
$\tau^{(s)}_{\mathrm{RR}}$ used in our simulation C3. The key to doing
this is to realize that Eqs.~\eqref{eq:crevol} and \eqref{eq:cxevol}
describe the coupled mode evolution in the physical case if we just
substitute $\tau^{(p)}_{\mathrm{RR}}$ for $\tau^{(s)}_{\mathrm{RR}}$.
The mode $X$ is capable of stopping the unstable growth of the
$r$-mode only when the magnitude of the second term on the right side
of Eq.~\eqref{eq:crevol} becomes comparable to the first.  Through an
analysis similar to the one that led to Eq.~\eqref{eq:inequalityb}, it is
straightforward to find the following condition on the saturation
amplitude of the $r$-mode,
\begin{equation}
\left[ \frac{P_R}{\tau^{(p)}_{\mathrm{RR}}} \right]^2 \simeq
\pi^2 \frac{|c_R^{\mathrm{sat}}|^2}{\gamma^{(p)}} \left| \frac{ \kappa^*_{XRR} }{ \epsilon_X }
\right|^2 \frac{\epsilon_X}{\epsilon_R}.
\end{equation}
We now use the upper limit for $|\kappa^*_{XRR}|$ from
Eq.~\eqref{eq:klimit} from our numerical evolution, to obtain a
\emph{lower} limit for the amplitude $c^{\mathrm{sat}}_R$ at which the
$r$-mode would be saturated in the physical case:
\begin{equation}
|c^{\mathrm{sat}}_R| \gg 20 \frac{P_R}{\tau^{(p)}_{\mathrm{RR}}}
\sqrt{\frac{\gamma^{(p)}}{\gamma^{(s)}}}.
\end{equation}
Since $\gamma^{(p)}>\gamma^{(s)}$ this equation yields $|c_R^{\mathrm{sat}}|\gg 4\times10^{-4}$ for run C3.
So if the dominant mode--mode coupling is of the form given
in Eqs.~\eqref{eq:crevol} and \eqref{eq:cxevol}, our simulation places a
relatively large lower limit on the $r$-mode saturation amplitude.
However, the $r$-mode could instead be limited by \emph{parametric
resonance} \cite{Dziembowski85} with a suitable pair of modes
(satisfying the resonance condition $\omega_R + \omega_Y + \omega_Z
\simeq 0$).  It appears that our simulation does not provide a very
strong lower limit on the saturation amplitude that could be imposed
by this kind of process.

\subsection{Dependence on the grid spacing}

\begin{figure}
\begin{center}
\includegraphics{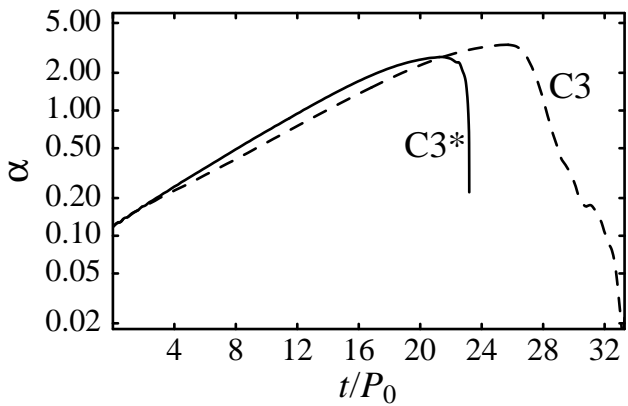} \caption{Numerical evolution of the
$r$-mode amplitude $\alpha$ in low-resolution run C3* (solid curve)
and in run C3 (dashed curve).
\label{fig:smallamp}}
\end{center}
\end{figure}

We wish to confirm that our standard computational grid can resolve
the spatial structure of the $r$-mode well enough to give reliable
predictions about the saturation amplitude of the mode.  For this
purpose, we have performed a simulation (run C3*) with the same
parameters of run C3, but on a grid with only half the spatial
resolution ({i.\ e.}, 32 cells in the radial direction, and 64 cells
in the axial and azimuthal directions).  Figure~\ref{fig:smallamp}
compares the evolution of $\alpha$ in runs C3 and C3*. The two curves
are very similar, but in run C3* saturation is reached a bit earlier,
at $t/P_0 = 21.4$, and at a somewhat lower amplitude $\alpha = 2.68$.
This may be caused by the larger numerical viscosity that must be
present in the coarser grid. The evolution of the other diagnostics is
also very similar in the two runs.

Thus, the simulation run (C3*) suggests that the qualitative results
of our simulations are independent of the resolution adopted. The
$r$-mode saturation amplitudes on the two grids agree to within about
20\%.  Interestingly, the extrapolation to the infinite resolution
case suggests that the physical saturation amplitude might be even
larger than 3.3.

\section{\label{sec:k1000} testing the saturation amplitude}

Even in the absence of a saturation mechanism due to mode--mode coupling
as described above, it is possible that the saturation amplitude in
our simulation might still depend on the strength of the
radiation-reaction force.  In our simulation we see that the $r$-mode
grows until density waves on the surface of the star break and form
shocks.  It is possible that this occurs just because in our
simulation we are pushing the fluid too hard with the
radiation-reaction force, much harder than it would be appropriate in the physical
case. To explore how the evolution depends on the strength of this
driving force, we go back to the time in run C3 before any signs of
nonlinear saturation are seen, when $\alpha = 1.8$.  We start a new
run (C4) there, increasing the value of $\kappa$ (which determines the
strength of the radiation-reaction force) to 5967 (1.33 times its
value in run C3). The new growth timescale is about $7.5 P_0$.
(Undoubtedly, a test with $\kappa \ll 4487$ would have been more
compelling; but our evolutions are so computationally expensive that
we were forced to increase rather than decrease the strength of the
driving force.)

In a separate run (C5), we test the influence of the \emph{history} of
the evolution of the $r$-mode on its saturation amplitude.  Namely, we
ask if an $r$-mode that started out as the {\it linear} initial data
of Eq.~\eqref{eq:initialmodel}, with a very large amplitude, would
evolve much differently from an $r$-mode that started out small and was
built up gradually to large amplitude by the radiation reaction
force. To answer this question, we start with the \emph{Fast}
equilibrium model, and we add a linear $r$-mode velocity field with
$\alpha_0 = 1.8$. For this run we keep $\kappa = 4497$.
\begin{figure}
\begin{center}
\includegraphics{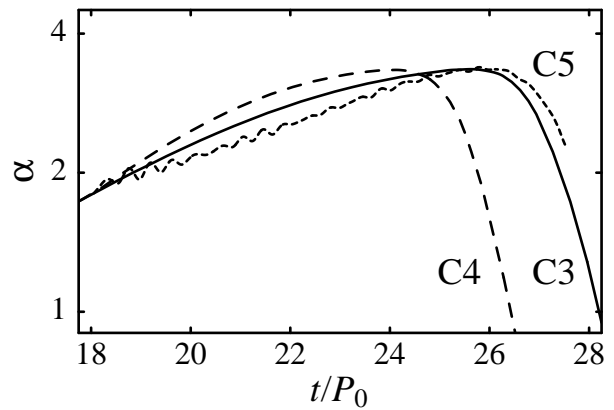} \caption{Numerical evolution
of the $r$-mode amplitude $\alpha$ in production runs C3--C5.
\label{fig:fastk1000alpha2amp}}
\end{center}
\end{figure}
\begin{figure}
\begin{center}
\includegraphics{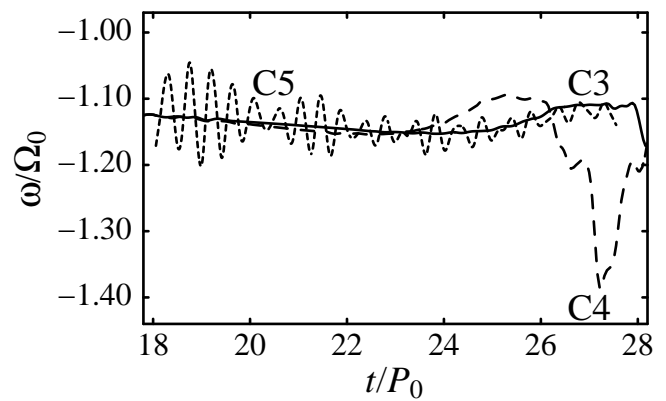} \caption{Numerical evolution
of the $r$-mode frequency $\omega$ in production runs C3--C5.
\label{fig:fastk1000alpha2omg}}
\end{center}
\end{figure}

\begin{figure}
\begin{center}
\includegraphics{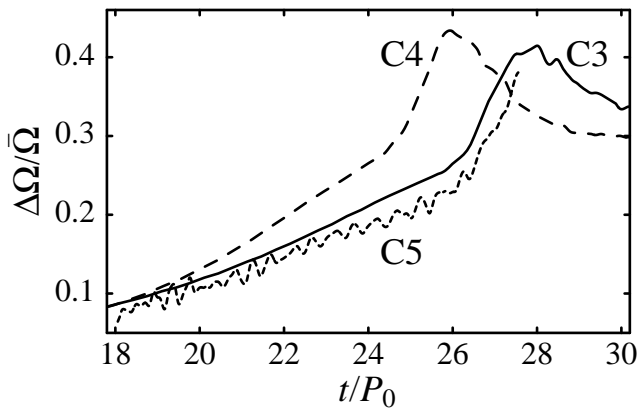} \caption{Numerical
evolution of the differential rotation $\Delta \Omega$ in
production runs C3--C5. \label{fig:fastk1000alpha2delomg}}
\end{center}
\end{figure}

Figures \ref{fig:fastk1000alpha2amp}, \ref{fig:fastk1000alpha2omg}, and
\ref{fig:fastk1000alpha2delomg} show the evolution of the diagnostic
parameters $\alpha$, $\omega$ and $\Delta \Omega$ for runs C3--C5. As
expected, the $r$-mode does grow faster in run C4, but its maximum
value is essentially the same (the maximum $\alpha = 3.338$ at $t =
24.12 P_0$) as in run C3. In this run, the $r$-mode amplitude
increases from $\alpha = 1.8$ to $\alpha = 3.338$ within a time
$\Delta t \simeq 6 P_0$ (compared to $\Delta t \simeq 8 P_0$ in run
C3) as would be expected given that the driving force is $\frac{4}{3}$
times that of run C3.

In run C5, the growth of the $r$-mode is initially slower than in run
C3, as the linear $r$-mode velocity field evolves toward its correct
nonlinear form.  Eventually its maximum occurs at essentially the same
amplitude as before ($\alpha = 3.337$).  Figures
\ref{fig:fastk1000alpha2amp} and \ref{fig:fastk1000alpha2omg} show
that during run C5 $\alpha$ and $\omega$ undergo short-period
oscillations; this happens because the initial velocity field is only
a small-amplitude approximation to the \emph{real} $m=2$ $r$-mode
eigenfunction. So other spurious modes with fairly large amplitude are
excited initially in run C5.  Note that these extraneous modes must
make nonzero contributions to $J_{22}$ if they are to show up in our
diagnostics. Here the extraneous modes cause a rapid modulation of
$\alpha$ and $\omega$ with a dominant period of about $0.5
P_0$. Finally, it is interesting to consider the evolution of $\Delta
\Omega$ (Fig.~\ref{fig:fastk1000alpha2delomg}), which is very similar
in the three runs.

These runs provide limited evidence that the saturation amplitude of
the $r$-mode does not depend (strongly) on our artificially large
radiation reaction force.  The nonlinear hydrodynamical process that
leads to shock formation appears to be triggered by attaining a
certain critical amplitude of the $r$-mode, with little dependence on
the strength of the radiation-reaction force.  Thus if no mode--mode
coupling occurs on timescales longer than our unphysically short
$\tau^{(s)}_{RR}$, then our results suggest that the maximum amplitude
$\alpha\approx3$ is a reasonable guess for the physical case ($\kappa
= 1$) as well.

\section{\label{sec:font}Free evolution}

Stergioulas and Font \cite{Stergioulas01} have also studied the
nonlinear evolution of $r$-mode initial data, but using relativistic
hydrodynamics in a fixed background geometry.  In their evolution
using this \emph{relativistic Cowling approximation}, the
gravitational interactions of the mode with itself and with the rest
of the star are neglected.  The principal difference between their
model and ours therefore is that theirs has no radiation reaction, and
no $r$-mode growth.

Stergioulas and Font find that, for an initial $r$-mode amplitude
$\alpha_0 = 1.0$, no \emph{significant} suppression of the mode is
observed during 13 rotation periods.  They define their mode amplitude
using a post-Newtonian expression for the eigenfunction that differs
from our Eq.~\eqref{eq:initialmodel} except in the Newtonian limit.
And their method of evaluating the mode amplitude numerically also
differs from ours. They read the mode amplitude from the value of the
fluid's velocity at a single point within the star, while we define
$\alpha$ in terms of integrals over the entire star.  In the
slow-rotation Newtonian limit our two definitions agree.  Stergioulas
and Font observe that the amplitude of the velocity oscillations
(shown in Fig.~2 of Ref.~\cite{Stergioulas01}) decrease by about 50\%
during the course of their simulation, an effect that they attribute
to numerical viscosity \cite{Stergioulas01}.  In order to compare our
own simulations more directly with theirs, we performed a series of
evolutions in which we turned off the radiation-reaction force by
setting $\kappa = 0$.

In production runs C6 and C7, we augment our rapidly rotating
equilibrium configuration with the approximate $r$-mode velocity field
of Eq.~\eqref{eq:initialmodel}. For run C6, we choose the initial
$\alpha_0$ so that $\alpha$ [as measured by our numerical diagnostic,
Eq.~\eqref{eq:alphadef}] is initially 1.8: the value at which we start
to observe deviations from exponential growth in run C3.  For run C7,
we choose $\alpha_0$ so that the initial $\alpha$ is 1.0, in order to
make a direct comparison with Stergioulas' and Font's published
results.  We have evolved these systems through respectively 11 and 7
initial rotation periods (several hydrodynamical timescales, according
to our rough estimate of the speed of sound for the rapidly rotating
model).
\begin{figure}
\begin{center}
\includegraphics{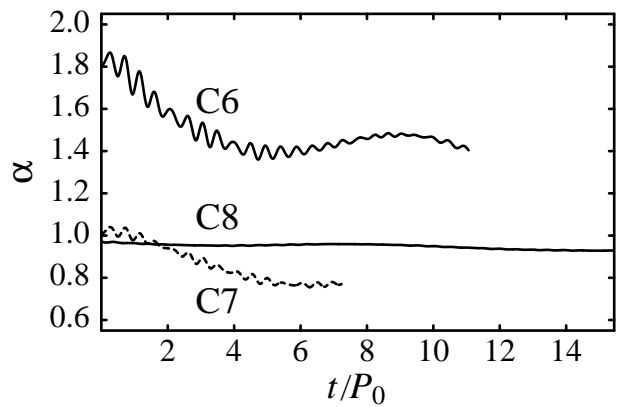} \caption{Numerical
evolution of the $r$-mode amplitude $\alpha$ in production runs
C6--C8. \label{fig:beta2gamma2delta2amp}}
\end{center}
\end{figure}

\begin{figure}
\begin{center}
\includegraphics{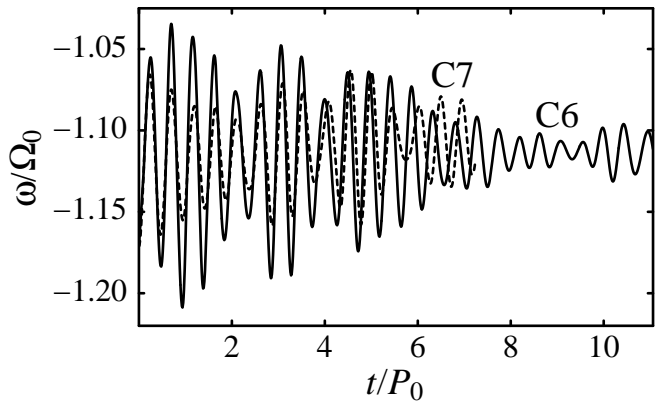} \caption{Numerical evolution of
the $r$-mode frequency $\omega$ in production runs C6 and C7.
\label{fig:beta2gamma2omg}}
\end{center}
\end{figure}

We plot the evolution of $\alpha$ and $\omega$ for these simulations
in Figs.~\ref{fig:beta2gamma2delta2amp} and
\ref{fig:beta2gamma2omg}. The wavy appearance of the curves suggests
that, by using the linear eigenfunction, Eq. \eqref{eq:initialmodel},
for amplitudes of order unity, we have excited spurious modes in
addition to the basic $m = 2$ $r$-mode. We have already observed this
behavior in run C5. The rapid modulation of $\alpha$ and $\omega$ has
a period of about $0.5 P_0$, and the amplitude of the modulation is
smaller for run C7. (This is reasonable: for lower $\alpha$ we expect
the approximate expression, Eq. \eqref{eq:initialmodel}, to be more accurate
and so to excite smaller amplitude spurious modes.)

\begin{figure}
\begin{center}
\includegraphics{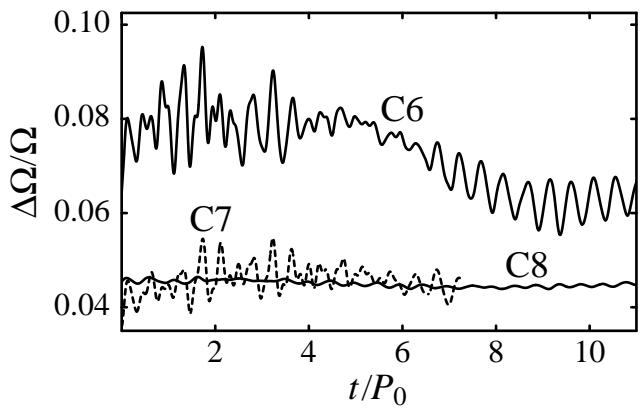} \caption{Numerical
evolution of the differential rotation $\Delta \Omega$ in
production runs C6--C8. \label{fig:beta2gamma2delta2delomg}}
\end{center}
\end{figure}

In both runs, $\alpha$ loses about 20\% of its initial value during
the first four rotation periods. In the next few rotation periods,
however, the average value of $\alpha$ remains unchanged (although in
run C6 we can see a further modulation of the amplitude with a period
of about $8 P_0$). Throughout the runs, the $r$-mode frequency
$\omega$ oscillates around $\omega = -1.12 \Omega_0$, consistent with
its value in run C3 for the same value of $\alpha$ ({i.\ e.},
1.44). As the run is started, the differential rotation $\Delta
\Omega$ (which is zero in the initial, rigidly rotating star)
increases almost immediately to values that are consistent with those
observed in run C3 for the same amplitude; compare
Figs.~\ref{fig:beta2gamma2delta2delomg} and \ref{fig:fastdelomg}.  As
$\alpha$ decreases, $\Delta \Omega$ decreases consistently.  (In run
C7, $\Delta \Omega$ settles to a value slightly higher than what we
expected from its value in run C3 when $\alpha = 0.82$; but we did not
run this evolution as far as run C6, so at the end of our simulation
the value of $\Delta \Omega$ might still be evolving.)

Finally, we study the free nonlinear evolution of an $r$-mode that was
\emph{grown} to the amplitude $\alpha = 1$. To do so, we go back to
the time in run C3 when $\alpha = 1$, and start a new run (C8) using
the C3 data at this time. We evolve these data setting $\kappa$ to
zero in the subsequent evolution. We follow this evolution through an
additional 15.4 initial rotation periods.  During this time the mode
amplitude $\alpha$ is essentially constant, see
Fig.~\ref{fig:beta2gamma2delta2amp}, except for a slow secular decline
due to numerical viscosity at 0.23\%/revolution, and a few very small
amplitude oscillations.  The $r$-mode frequency is quite constant, and
the phase coherence function, and the differential rotation $\Delta
\Omega$ also remain quite small in this case (see
Fig.~\ref{fig:beta2gamma2delta2delomg}).  The $r$-mode amplitude in
run C3 remains above unity for 14.3 rotation periods, so run C8
demonstrates that the LSU hydrodynamic code~\cite{Tohline97,Tohline01}
used here reliably and stably evolves large amplitude $r$-modes in
rapidly rotating stars for the duration of our simulations.

Comparing runs C6, C7, and C8, we infer that the strong decrease in
the amplitude observed in runs C6 and C7 occurs as nonlinear
hydrodynamics reorganizes the initial linear $r$-mode velocity field
to the correct nonlinear form for amplitudes of order unity. After the
reorganization is complete (within a few rotation periods), $\alpha$
decreases only because of numerical viscosity. (In run C5, this same
phenomenon caused the slower growth of the amplitude compared to run
C3.)  By contrast, the small decrease in run C8 appears to be caused
entirely by numerical viscosity.

Altogether, we find that our results are compatible with those of
Stergioulas and Font \cite{Stergioulas01}: no nonlinear saturation
effect is evident in the free nonlinear evolution of $r$-modes, at
least for amplitudes of order unity. 

\section{\label{sec:k1000b}repeated spindown episodes?}

The first attempt to analyze the nonlinear evolution of $r$-modes by
Owen, \emph{et al.}\ \cite{Owen98} was based on a simple two-parameter
model consisting of a rotating star with angular velocity $\Omega$ and
its $r$-mode with amplitude $\alpha$.  Using this model the mode was
found to grow exponentially until it reached some maximum level
$\alpha_{\mathrm{max}}$, where it was assumed to remain saturated.
Energy and angular momentum were expected to be removed from the star
by gravitational radiation during this saturation phase until the
$r$-mode regained stability (because of increased internal dissipation
brought about by cooling or because the angular momentum of the star
was reduced to a very low level).  In this initial picture
gravitational radiation was expected to spin down the star on a
timescale of about one year.  The radiation emitted was expected to
sweep down in frequency from $\frac{4}{3}$ times the initial angular
velocity of the star to $\frac{4}{3}$ times its final value: ranging
from perhaps $1$ kHz initially to perhaps $100$ Hz.

Our simulations suggest a very different picture.  We find that, once
the amplitude of the $r$-mode reaches $\alpha_{\mathrm{max}}$, it is
quickly reduced by the action of the breaking waves and shocks,
instead of remaining saturated at this value for a very long time.  At
the end of our simulation the star still has 60\% of its initial
angular momentum, and its average angular velocity is 77.5\% of
$\Omega_0$. Thus the star is left rotating relatively rapidly, leaving
open the possibility of subsequent episodes of $r$-mode instability
and spindown.

\begin{figure}
\begin{center}
\includegraphics{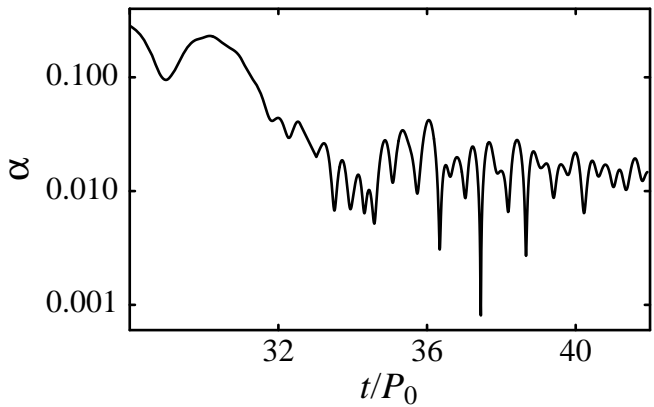} \caption{Numerical evolution of the
$r$-mode amplitude $\alpha$ in the extended run C4.
\label{fig:k1000amp}}
\end{center}
\end{figure}

\begin{figure}
\begin{center}
\includegraphics{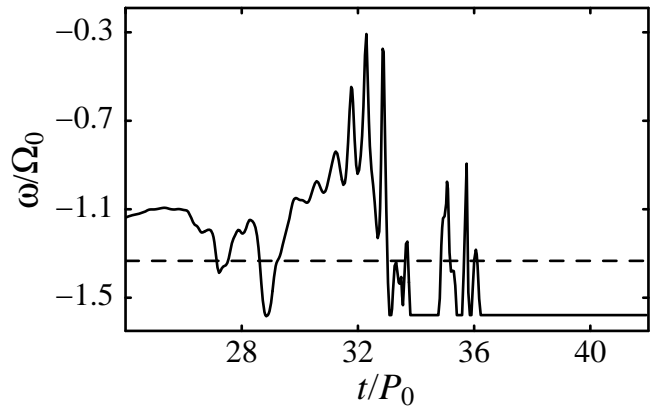} \caption{Numerical evolution of the
$r$-mode frequency $\omega$ in the extended run C4.
\label{fig:k1000omg}}
\end{center}
\end{figure}

To investigate this possibility, we extend run C4, evolving our star
for 13 more initial rotation periods after $\alpha$ has gone back to
its initial value (0.1), or (equivalently) for nine periods after
$\alpha$ reaches its minimum ($\sim 0.01$). The evolution of the
amplitude for this case is plotted in Fig.~\ref{fig:k1000amp}.  After
$t = 33 P_0$, the fluid motion is quite turbulent, but we see no sign
that $\alpha$ is starting to grow again. The evolution of the $r$-mode
frequency (Fig.~\ref{fig:k1000omg}) is also erratic, probably because
here the sinusoidal approximation begins to fail (remember that
$\omega$ is approximated as $-(1/J_{22})d|J_{22}|/dt$). In fact, after
$t = 33 P_0$ we have found it necessary to impose an \emph{ad hoc}
limit on the value of $\omega$; otherwise $\omega$ grows to about $-17
\Omega_0$, and the radiation-reaction force (proportional to
$\omega^6$) becomes huge, pushing the fluid to superluminal
velocities.

Nine periods should be more than enough to see a second $r$-mode
growth episode, if it occurs at all.  Although at the end of the
simulation the average angular velocity of the star is lower than
$\Omega_0$, the growth timescale is determined by the $r$-mode
frequency, which is even higher than at the beginning of the run.
What keeps the $r$-mode then from resuming its growth?

\begin{figure}
\begin{center}
\includegraphics{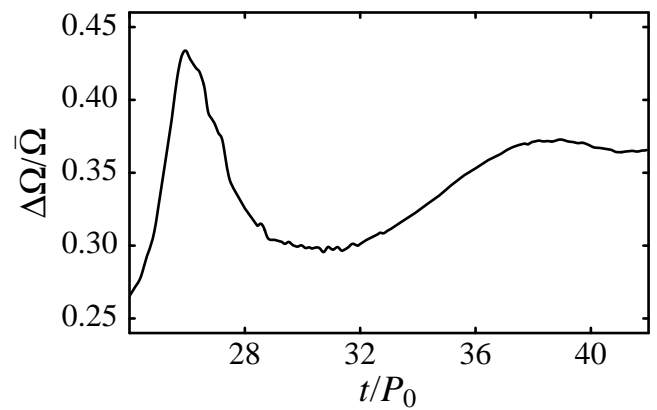}
\caption{Differential rotation $\Delta \Omega$ through the extended run C4.
\label{fig:k1000delomg}}
\end{center}
\end{figure}

\begin{figure}
\begin{center}
\includegraphics[width=3.4in]{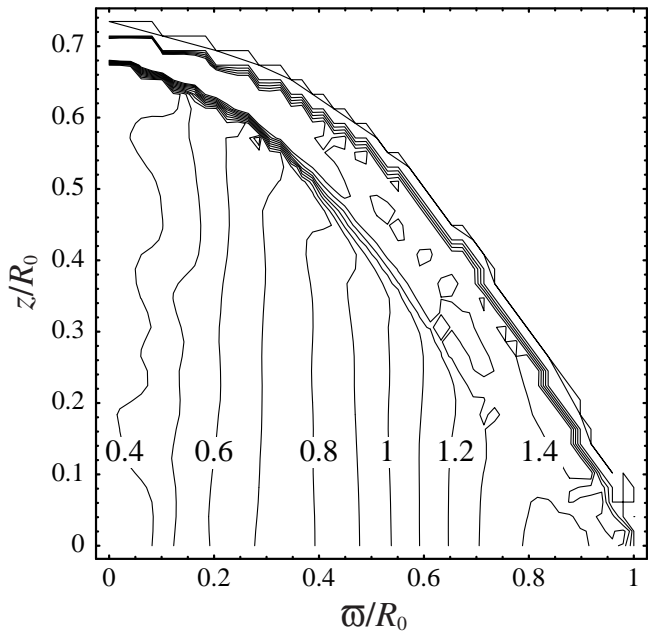} \caption{Meridional
structure of differential rotation at the end of production run
C4. This contour plot shows level contours for the value of the
azimuthally averaged angular velocity
$\Omega(\varpi,z)/\bar{\Omega}$, at time $t = 42 P_0$.
\label{fig:fastdelstructb}}
\end{center}
\end{figure}

One hypothesis is that because of its strong differential rotation the
post-spindown configuration of the star is one which stabilizes the
$r$-mode. The value of $\Delta \Omega$ for the last few periods is
plotted in Fig.~\ref{fig:k1000delomg}. The increase of $\Delta \Omega$
observed between $t = 32 P_0$ and $t = 36 P_0$ is not caused by
radiation reaction, but by a global, energy-conservative
reorganization of the fluid. At the end of this process, the spatial
structure of differential rotation is very different from what it was
at $\alpha_{\mathrm{max}}$: compare Fig.~\ref{fig:fastdelstruct} ($t =
25.6 P_0$ in run C3) with Fig.~\ref{fig:fastdelstructb} ($t = 42 P_0$
in run C4). The latter plot shows a star that is rotating on cylinders
(except for the outer layer), with $\Omega(\varpi,z)$ almost
proportional to $\varpi$.

Karino, \emph{et al.}\ \cite{Karino00} derived linearized structure
equations for the $r$-modes of differentially rotating Newtonian
stars. When differential rotation is so strong that \emph{corotation
points} appear (that is, when there exists a $\varpi$ such that
$\omega + m \Omega(\varpi) = 0$), the mode equations go singular. (The
presence of a corotation point at the cylindrical radius $\varpi$
means that the velocity pattern of the mode appears to stand still in
the frame rotating with angular velocity $\Omega(\varpi)$.) A
comparison of the differential rotation of
Fig.~\ref{fig:fastdelstructb} with the value of $\omega$ suggests the
presence of corotation points in the final configuration of our
star. By itself, however, the singularity of the linearized mode
equations does not necessarily mean that $r$-modes are impossible.

A second, probably more likely possibility is that, in the very noisy
environment manifest in Figs.~\ref{fig:k1000amp} and
Figs.~\ref{fig:k1000omg}, the growing $r$-mode is unable to get locked
in phase with the approximate expression for the driving force that we
use here.  The actual radiation reaction force
[Eq.~\eqref{eq:rrforce}] is a function of the frequency of the
$r$-mode. Since we do not know exactly what this frequency is, we use
the expression \eqref{eq:genomega} to approximate it.  This
approximation works extremely well as long as the $r$-mode makes the
dominant contribution to $J^{(1)}_{22}$; yet, in the turbulent
post-spindown environment, the $r$-mode no longer dominates the
evolution of $J_{22}$. Hence, our expression for the gravitational
radiation reaction force is no longer correct: it fails to maintain
phase coherence with the $r$-mode and so prevents the growth of the
mode.

If the $r$-mode really does not exist in the chaotic post-spindown
environment, then it will be necessary to wait for viscosity to damp
differential rotation before the $r$-mode can grow again.  However,
viscosity might be unable to do this before the star cools so much
that the $r$-mode is stabilized (either because the star forms a crust
or because viscosity itself has grown too strong). This possibility is
worrisome, because the same environmental conditions (strong
differential rotation and generalized noise) that characterize the end
of run C4 are likely to occur in the young supernova remnants where
$r$-modes are expected to arise in nature.  Still, we think it more
likely that the absence of a second growth episode in our simulation
is the result of our expression for the radiation reaction force,
which is too simple for this chaotic situation.

\section{\label{conclusion}Conclusions}

We have completed a series of numerical 3D hydrodynamical simulations
of the nonlinear evolution of the GR driven instability in the
$r$-modes of rotating neutron stars.  We have verified that the
current-quadrupole GR reaction force implemented in our code is
accurate by reproducing the analytical predictions (for slowly
rotating stars) with our full 3D numerical integration code.  In
our simulations, the amplitude of the ($m=2$) $r$-mode is driven to a
value of about three before nonlinear hydrodynamic forces stop its
growth by the formation of shocks and breaking surface waves.  We
showed that the value of this maximum amplitude is insensitive to the
strength of the GR driving force by repeating the simulation for
different strengths and different initial fluid configurations.  We
also repeated our simulation using a coarser numerical grid to verify
the robustness of our results (the maximum mode amplitude changes only
by about 20\% when the number of grid points is reduced by a factor of
8), and to show in particular that numerical viscosity is not playing a
critical role in our simulations.

In our simulation we have artificially increased the strength of the GR
reaction force in order to reduce the problem to one that can be
studied with the available computer resources.  We have shown,
however, that the results of our simulation can be used to infer
limits on the real physical problem as well.  We used the results of
our simulations to derive a lower limit of a few times $10^{-4}$ on
the saturation amplitude of the $r$-mode in a real neutron star due to
possible (but unseen) nonlinear mode--mode couplings.  This lower limit
applies to couplings with modes that are well described by our
simulation: that is, the modes of a barotropic fluid with spatial
structures larger than about 2\% of the radius of the star.

Recent analysis of the effects of magnetic fields~\cite{Rezzolla01}, and exotic forms of bulk viscosity~\cite{Lindblom01b} suggest that the $r$-mode instability may not play as important a role in astrophysical situations as was once thought.  However, the considerable uncertainty that exists about both the macroscopic and microscopic states of a neutron star makes it impossible at the present time to conclude that the $r$-mode instability plays no astrophysical role.  Thus it seems reasonable to us that some effort be put into gravitational wave searches for $r$-mode signals having forms qualitatively similar to those predicted by simulations such as this.

\begin{acknowledgments}
We thank L.~Bildsten, J.~Friedman, Yu.~Levin, B.~Owen, N.~Stergioulas,
K.~Thorne, G.~Ushomirsky, and R.~Wagoner for helpful discussions. We
also thank H.~Cohl, J.~Cazes, and especially P.~Motl for contributions
to the LSU hydrodynamic code.  This research was supported by NSF
grants PHY-9796079, AST-9987344, AST-9731698, PHY-9900776,
PHY-9907949, and PHY-0099568 and NASA grants NAG5-4093, NAG5-8497 and
NAG5-10707. We thank NRAC for computing time on NPACI facilities at
SDSC where tests were conducted; and we thank CACR for access to the
HP V2500 computers at Caltech, where the primary simulations were
performed.
\end{acknowledgments}

\appendix*

\section{\label{cylindrical}Useful expressions in cylindrical coordinates}

In this Appendix we give explicit expressions in cylindrical
coordinates $(\varpi, z, \varphi)$ for a number of useful
quantities used in our simulations.  The components of the
the initial $r$-mode velocity field used in our numerical
evolutions are
\begin{equation}
v^\varpi =
\alpha_0 \sqrt{\frac{5}{16\pi}} \frac{\Omega_0}{R} z \varpi \sin 2\varphi,
\label{eq:rmodevr}
\end{equation}
\begin{equation}
v^z = -\alpha_0 \sqrt{\frac{5}{16\pi}} \frac{\Omega_0}{R} \varpi^2 \sin 2\varphi,
\label{eq:rmodevz}
\end{equation}
and
\begin{equation}
v^{\hat{\varphi}} = \Omega_0 \varpi
+ \alpha_0 \sqrt{\frac{5}{16\pi}} \frac{\Omega_0}{R} z \varpi \cos 2\varphi.
\label{eq:rmodevphi}
\end{equation}
We refer the azimuthal component of the velocity to the
orthonormal coordinate $\hat{\varphi}$, so that
$v^{\hat{\varphi}}$ and $v_{\hat{\varphi}}$ have the same
numerical value and we can use them interchangeably.

The integrals that determine $J_{22}$ and its first
time-derivative $J^{(1)}_{22}$ are,
\begin{equation}
J_{22} = \sqrt{\frac{5}{16\pi}} \int \rho e^{-2 i \varphi}
[z v_{\hat{\varphi}} + i (z v_\varpi - \varpi v_z) ]
\varpi^2 d \varpi d z d \varphi,
\end{equation}
and
\begin{equation}
J^{(1)}_{22} = \sqrt{\frac{5}{16\pi}} \int \rho e^{-2 i \varphi}
[T_1 + i T_2]
\varpi d \varpi dz d \varphi,
\end{equation}
where
\begin{eqnarray}
T_1 & \equiv & 2 z v_\varpi v_{\hat{\varphi}} - \varpi v_z
v_{\hat{\varphi}}
- z \frac{\partial \Phi}{\partial \varphi}, \\
T_2 & \equiv & z (v^2_\varpi - v^2_{\hat{\varphi}}) - \varpi
v_\varpi v_z + \varpi^2 \frac{\partial \Phi}{\partial z} - z
\varpi \frac{\partial \Phi}{\partial \varpi}.
\end{eqnarray}

The components of the radiation-reaction force in cylindrical
coordinates are obtained from Eq.~(\ref{eq:rrforce}) by expressing
the current multipole tensor $S_{jk}$ in terms of the current
multipole moments $J_{2m}$ via
Eqs.~\eqref{eq:sandja}--\eqref{eq:sandjc}:
\begin{eqnarray}
F^{\mathrm{GR}}_z &=&
- \kappa \frac{16}{45} \sqrt{\frac{4 \pi}{5}} \frac{G}{c^7}
\rho \varpi \\
&& \times {\mathrm{Im}} \biggl\{ e^{2 i \varphi}
\Bigl[ 3 (v_\varpi + i v_{\hat{\varphi}}) J^{(5)}_{22} + \varpi J^{(6)}_{22} \Bigr] \biggr\}, \nonumber
\end{eqnarray}
and
\begin{equation}
F^{\mathrm{GR}}_{\hat{\varphi}} - i F^{\mathrm{GR}}_\varpi =
\kappa\frac{16}{45} \sqrt{\frac{4 \pi}{5}} \frac{G}{c^7}
\rho \varpi \, e^{2 i \varphi}
\Bigl[ 3 v_z J^{(5)}_{22} + z J^{(6)}_{22} \Bigr],
\end{equation}
where $\kappa=1$ in general relativity theory.  The fifth and sixth
time derivatives of $J_{22}$ are obtained as $J^{(5)}_{22} = \omega^4
J^{(1)}_{22}$, and $J^{(6)}_{22} = - \omega^6 J_{22}$.

\end{document}